\documentclass[aps,prd,amsmath,amsfonts,amssymb,twocolumn,nofootinbib]{revtex4}

\usepackage{amssymb}
\usepackage{graphicx,psfrag,color}
\usepackage{dcolumn}
\usepackage{bm}
\usepackage{mathbbol}
\usepackage{slashed}
\usepackage[utf8]{inputenc}
\DeclareUnicodeCharacter{2009}{\,} 

\begin{document}
\flushbottom

\title{Asymmetric particle-antiparticle Dirac equation: first quantization}

\author{Gustavo Rigolin}
\email{rigolin@ufscar.br}
\affiliation{Departamento de F\'isica, Universidade Federal de
S\~ao Carlos, 13565-905, S\~ao Carlos, S\~ao Paulo, Brazil}

\date{\today}

\begin{abstract} 
We derive a Dirac-like equation, the \textit{asymmetric Dirac equation}, where particles and antiparticles sharing the same wave number have different energies 
and momenta. We show that this equation is Lorentz covariant under proper Lorentz 
transformations (boosts and spatial rotations) and also determine the corresponding transformation law for its wave function. We obtain a formal connection between 
the asymmetric Dirac equation and the standard Dirac equation and we show that by 
properly adjusting the free parameters of the present wave equation we can make it reproduce
the predictions of the usual Dirac equation. We show that the rest mass of 
a particle in the theoretical framework of the asymmetric Dirac 
equation is a function of a set of four parameters, which are 
relativistic invariants under proper Lorentz transformations. These four parameters 
are the analog to 
the mass that appears in the standard Dirac equation. 
We prove that in order to guarantee the covariance of the asymmetric Dirac equation under
parity and time reversal operations (improper Lorentz transformations) as well as under 
the charge conjugation operation, these four parameters change sign in exactly the same 
way as the four components of a four-vector. 
The mass, though, being a function of the square of those parameters 
remains an invariant. We also extensively study the 
free particle plane wave solutions to the asymmetric Dirac equation and derive its  
energy, helicity, and spin projection operators as well as several Gordon's identities.
The hydrogen atom is solved in the present context after applying the minimal coupling 
prescription to the asymmetric Dirac equation, which also allows us to appropriately 
obtain its non-relativistic limit. 
\end{abstract}


\maketitle

\section{Introduction} 

The wave function $\psi(x)$ that solves the standard non-relati\-vis\-tic 
Schr\"odin\-ger equation \cite{sch26} transforms after a Galilean boost 
as \cite{bal98}
\begin{equation}
\psi(x) = e^{\frac{i}{\hbar}\theta(x')} \psi'(x'). 
\label{psiStransform}
\end{equation}
The phase $\theta(x')$ is a function of $x' = (ct',\mathbf{r'})$ and of the relative velocity
$\mathbf{v}$ between the two inertial reference frames $S$ and $S'$,
\begin{equation}
\theta(x') = \frac{mv^2}{2}t' + m \mathbf{v\cdot\mathbf r'} + cte.
\label{thetaS}
\end{equation}
Here $cte$ is a real constant (usually set to zero \cite{bal98}), $v^2=\mathbf{v\cdot v}$,
$m$ is the mass of the particle, $t'$ and $\mathbf{r'}$ are respectively the time
and position of the particle in $S'$, $\hbar$ is Planck's constant divided by $2\pi$,
and $c$ is the speed of light.
We must have $\psi(x)$ transforming as given by Eq.~(\ref{psiStransform}) if we 
want the Schr\"odinger equation to be covariant under a Galilean boost \cite{bal98}. 

The transformation law of the Schr\"odinger wave function shows 
that we can have a logically consistent theory where a complex `scalar' field obeys a more general transformation law under a symmetry operation.
In other words, rather than assuming that the complex field $\psi(x)$ 
is a strict scalar, we can relax this assumption and demand only that the bilinear $\psi(x) \psi^*(x)=|\psi(x)|^2$ be a scalar under a 
given symmetry operation ($|\psi(x)|^2=|\psi'(x')|^2$). 

Is it possible to extend the transformation rule given by Eqs.~(\ref{psiStransform}) and (\ref{thetaS}) in a consistent way to the relativistic domain? 
What is then the relativistic wave 
equation covariant under a proper Lorentz transformation (boosts and spatial rotations) if the wave function transforms now according to this relativistic extension? In Ref. \cite{rig22} we answered in the affirmative the first
question above and derived the most general relativistic wave equation covariant under  proper Lo\-ren\-tz transformations compatible with the relativistic extension of the transformation rule given
by Eqs.~(\ref{psiStransform}) and (\ref{thetaS}). 

It turned out that the relativistic wave equation obtained in Ref. \cite{rig22}, which we
called the Lorentz covariant Schr\"o\-din\-ger equation, has both first and second order 
time and space derivatives. The main goal of this work is to obtain a consistent
relativistic wave equation that has at most first order time and space derivatives 
and that is, at the same time, compatible with the dispersion relations for particles and antiparticles that naturally emerge when working with the Lo\-ren\-tz covariant 
Schr\"o\-din\-ger equation \cite{rig22}.
We also show how the wave function of this first order differential equation transforms
under a proper Lorentz transformation and we prove 
that the wave equation is covariant under those transformations.

As it will become clear in the following pages, the first order 
wave equation we obtain is a Dirac-like spinorial equation \cite{dir67,gre00,man86,gre95} 
and its connection to the 
standard Dirac equation will also be given. Throughout our formal developments we will
try to build a spinorial wave equation that is as close as possible to the standard Dirac equation, pointing along the way the main differences between those two equations. And 
since the main difference between them is the fact that the wave equation here derived leads to relativistic energies for particles and antiparticles that are no longer degenerate, we
will from now on call it \textit{asymmetric Dirac equation}.

We should also mention that the physical motivation underlying the mathematical ideas and techniques of the present work stems from the fact that almost
all observables in a quantum field theory are bilinear functions of the fields.\footnote{For instance, the energy, the momentum, the angular momentum, and the charge are all functions of bilinears, i.e., functions
of quantities depending both on $\Psi(x)$ and $\Psi^\dagger(x)$. See 
Eqs. (\ref{ham}), (\ref{p}), (\ref{jvec}), and (\ref{Q}). See also
the sequel to this work, Ref. \cite{rig23}, for a complete discussion 
about the Dirac bilinears.} 
Therefore, a spinorial quantum field theory having a more general transformation rule, akin to what we have for the Lorentz covariant Schr\"odinger fields \cite{rig22},
should lead to a consistent theory compatible with all known experimental facts.
This can be accomplished if the bilinears transform in exactly the same way as the standard Dirac bilinears do after a given symmetry operation. We only need the 
bilinears, not the fields themselves, to transform as usual in order to recover
the predictions of the standard Dirac theory.

In the last part of this work we present some further formal developments, paving the 
way to the second quantization of the asymmetric Dirac equation that will be presented
elsewhere \cite{rig23}, and we apply the 
asymmetric Dirac equation in several interesting scenarios.
We start by first obtaining the free particle plane wave solutions of the 
asymmetric Dirac equation, which allows us to 
build its energy, helicity, and spin projection operators as well as derive several 
Gordon's identities. 
We then introduce electromagnetic interactions via the minimal coupling prescription. 
This allows us to obtain the non-relativistic limit of the asymmetric Dirac equation and to model the hydrogen atom using
the asymmetric Dirac equation. 
We also study how 
the asymmetric Dirac equation responds to the parity, time reversal, and charge 
conjugation symmetry operations. Finally, we show the Lagrangian density that leads to
the asymmetric Dirac equation and we derive from it the most important conserved Noether currents.

\section{The Lorentz covariant Schr\"odinger equation}
\label{secII}

Before we start the derivation of the asymmetric Dirac equation, it is important first to
present the Lorentz covariant Schr\"odinger equation and its main features needed for
our subsequent analysis. This will also help us 
set the notation and most of the terminology that will be used throughout this work. 

A careful investigation of the meaning of the first two terms in the right hand side of 
Eq.~(\ref{thetaS}), carried out in ref. \cite{rig22}, showed that for a free particle $mv^2/2$ and $m\mathbf{v}$ are, respectively, the kinetic energy and momentum ``gained'' 
by the particle of mass $m$ when we solve the Schr\"odinger equation in the reference frame $S$ instead of $S'$, with $S'$ moving away from $S$ with velocity $\mathbf{v}$. With this 
understanding, we postulated that in the relativistic regime we have after a 
boost \cite{rig22}
\begin{equation}
\psi(x) = e^{\frac{i}{\hbar}\theta(x')} \psi'(x')
\label{psiLtransform}
\end{equation}
and
\begin{equation}
\theta(x') = (\gamma - 1)mc^2 t' + \gamma m \mathbf{v\cdot\mathbf r'} + cte,
\label{thetaL}
\end{equation}
with $\gamma = 1/\sqrt{1-v^2/c^2}$ the Lorentz factor. Here $(\gamma - 1)mc^2$ and 
$\gamma m \mathbf{v}$ are, respectively, the ``gained'' relativistic kinetic energy and the relativistic momentum for a particle with rest mass $m$ when we describe the particle
in reference frame $S$ instead of $S'$. If in $S'$ the particle is at rest, 
$(\gamma - 1)mc^2$ and $\gamma m \mathbf{v}$ are, respectively,  the particle's 
relativistic kinetic energy and relativistic momentum from the point of view of $S$.  

With this transformation law for $\psi(x)$ we searched for the wave equation whose 
wave function transforms according to it and that is covariant under proper Lorentz transformations. With the aid of three extra reasonable assumptions that we list below,
we obtained the following free particle wave equation,
\begin{equation}
\frac{1}{c^2}\frac{\partial^2\psi(x)}{\partial t^2} - \nabla^2 \psi 
- i\frac{2m}{\hbar}\frac{\partial \psi(x)}{\partial t}=0.
\label{lcse}
\end{equation}

The extra three assumptions that together with Eqs.~(\ref{psiLtransform}) and 
(\ref{thetaL}) led uniquely to the wave equation (\ref{lcse}) were:
\begin{enumerate}
\item[(1)] The non-relativistic limit of the wave equation we are looking for
should be the 
Schr\"odinger equation.\\
\item[(2)] The wave equation should be isotropic, namely, covariant under three-dimensional spatial rotations in the same sense as the non-relativistic Schr\"odinger equation is. In 
other words, after a spatial rotation and assuming $\psi(x) = \psi'(x')$, we must 
get the same wave equation.\\
\item[(3)]  The wave equation should be a homogeneous linear partial differential 
equation of order not greater than two and with constant coefficients multiplying the derivatives.
\end{enumerate}

Looking at Eq.~(\ref{lcse}) we realize that there is no first order spatial derivatives. 
This lack of symmetry between the time derivative and the spatial derivatives is a 
consequence of assumption (2) listed above. In order to remedy that, and get a wave
equation fully symmetric in first and second order derivatives, we removed assumption (2)
and postulated that for any proper Lorentz transformation (boosts or spatial rotations) 
the wave function should transform as given in Eq.~(\ref{psiLtransform}), 
with $\theta(x')$ being a linear function of the space-time coordinates 
$x^{0'}=ct',x^{1'}=x',x^{2'}=y',x^{3'}=z'$. 

In particular, for an infinitesimal proper Lorentz transformation,  
\begin{equation}
x^\mu = x^{\mu'} - \epsilon^{\mu\nu}x_{\nu'}, 
\label{lt}
\end{equation}
we have  
\begin{equation}
\theta(x') = -i\epsilon_{\mu\nu}\kappa^{\mu}x^{\nu'},
\label{thetaInf}
\end{equation}
where $\epsilon^{\mu\nu}$ is the infinitesimal antisymmetric tensor related to the 
proper Lorentz transformation being implemented \cite{gre00,man86,gre95}. The four 
\textit{real} parameters $\kappa^0, \kappa^1,
\kappa^2$, and $\kappa^3$ will
be defined in a moment but the important point that we should stress now is the fact
that \textit{$\kappa^\mu$ is not a four-vector}. They are four relativistic invariants 
of the present theory which are related to the rest mass of the particle \cite{rig22}. 

Therefore, if the wave function transforms according to Eqs.~(\ref{psiLtransform}) and (\ref{thetaInf}), the following wave equation is covariant under proper Lorentz transformations,
\begin{equation}
\partial_\mu\partial^\mu\psi(x) - 2i\kappa^\mu\partial_\mu\psi(x) = 0.
\label{we}
\end{equation}
Here the metric is $g_{\mu\nu}=\mbox{diag}\{1,-1,-1,-1\}$, 
a covariant vector is given by $x_\mu=g_{\mu\nu}x^\nu$, 
and the covariant four-gradient is 
$
\partial_\mu = 
\left(\frac{\partial}{\partial x^0},\frac{\partial}{\partial x^1},\frac{\partial}{\partial x^2},\frac{\partial}{\partial x^3}
\right)$. It is also implicit the Einstein summation convention, with Latin indexes running from $1$ to $3$ and Greek ones from $0$ to $3$, and $g^{\mu\nu}=g_{\mu\nu}$
since we are working with a Minkowski spacetime. 

If we insert the following ansatz into Eq.~(\ref{we}),
\begin{equation}
\psi(x) = e^{i\kappa x}\phi(x)=e^{i\kappa_\mu x^\mu}\phi(x),
\label{GLStoKG}
\end{equation}
we get
\begin{equation}
\partial_\mu\partial^\mu\phi(x) + \kappa^2\phi(x) = 0.
\label{weKG}
\end{equation}

Equation (\ref{weKG}) can be identified with the Klein-Gordon one if we set
\begin{equation}
\kappa^2=\kappa_\mu\kappa^\mu = \mu^2=\widetilde{m}^2=(mc/\hbar)^2,
\label{dispKG}
\end{equation}
with $m$ being the rest mass of the scalar particle described by the Klein-Gordon equation.

Thus, we identify the rest mass of our particle as
\begin{equation}
m  = \frac{\hbar\sqrt{\kappa^2}}{c}.
\label{restmass}
\end{equation}
With this identification we can show that whenever we have self interactions that respect
the Lorentz symmetry or electromagnetic interactions modeled via the minimal coupling
prescription, the Lorentz covariant Schr\"odinger equation and the Klein-Gordon
one lead to the same predictions \cite{rig22}.
Note that Eq. (\ref{lcse}) is a particular case of (\ref{we}) since we can get the former from the latter by setting $\kappa^0=mc/\hbar$ and $\kappa^j=0$. Also, if we assume
no preferred orientation we must have $\kappa^1=\kappa^2=\kappa^3$.

Furthermore, Eq.~(\ref{restmass}) tells us that what we identify as the rest mass of 
a particle has its origin from essentially two parts. A ``time-like''  
contribution coming from $\kappa^0$ and a ``space-like'' one coming from $\kappa^j$.
It is the square root of $(\kappa^0)^2-|\bm{\kappa}|^2$, where 
$\bm{\kappa}=(\kappa^1,\kappa^2,\kappa^3)$, 
that is proportional to the mass of the particle. To avoid an imaginary mass 
and properly relate the Lorentz covariant Schr\"odinger equation to 
the Klein-Gordon one we need $|\kappa^0|>|\bm{\kappa}|$. Apart from that, we are free
to set any value we wish to $\kappa^\mu$ as long as we guarantee the validity of
Eq.~(\ref{restmass}). The full implications of this freedom to choose $\kappa^\mu$ are 
not yet completely understood.
However, as highlighted in Ref. \cite{rig22}, it may help us model condensed
matter systems that are spatially anisotropic \cite{saf93,zha19} 
or it might shed a different light in
our understanding of rest mass \cite{stu41,fol50,tuc99,pol01,cav12,con15,kha17} and mass 
renormalization procedures.   

We should also highlight that when we second quantize
Eq.~(\ref{we}), particles and antiparticles with the same rest mass no longer have 
degenerate energies \cite{rig22}. We either have $E_\mathbf{p}^+=-\hbar c \kappa^0 + 
E_\mathbf{p}$ or $E_\mathbf{p}^-=\hbar c \kappa^0 + E_\mathbf{p}$, where
\begin{displaymath}
E_\mathbf{p} = \sqrt{m^2c^4 + |\mathbf{p}|^2c^2} 
\end{displaymath}
is the standard relativistic energy.
The $+$ and $-$ sign in $E_\mathbf{p}^\pm$ remind us that $E_\mathbf{p}^+$ comes 
from the positive energy solutions and  $E_\mathbf{p}^-$ from the negative energy 
solutions of the Lorentz covariant Schr\"odinger equation.
The same feature is ob\-ser\-ved for the momentum, where we either have $p^j_+=-\hbar \kappa^j 
+ \hbar k^j$ or $p^j_-=+\hbar \kappa^j + \hbar k^j$, with $k^j$ being the particle or 
antiparticle wave number. The interpretation and physical significance of this ``rest momentum'' is still an open problem \cite{rig22}. See also 
Refs. \cite{din04,ber97,col98,car06,ces15,sak17,edw18,and04} on how to
generate an asymmetry between matter and antimatter via Lorentz-violating theories
and Refs. \cite{kos89,kos95,ber07,ber07b,ber08a,ber08,rol13,rol15,rol17} on further strategies to build Lorentz-violating theories.

\section{Obtaining the asymmetric Dirac equation}

We can better appreciate and understand the techniques used to obtain the 
asymmetric Dirac equation if we first review in a modern notation the 
path taken by Dirac himself to get to his equation \cite{dir67}. After that it 
will become clearer why we have to follow a slightly different route  to obtain the 
asymmetric Dirac equation than the one used
by Dirac.

\subsection{The Dirac equation}
\label{de}

Dirac wanted to get a first order differential equation giving the right energy-momentum relation for a relativistic free particle. As such, each component of the spinor $\Psi_D(x)$
had to satisfy the Klein-Gordon equation \cite{dir67,gre00}. Note that we are using 
the subscript ``D'' to distinguish the standard Dirac spinor from the spinor 
$\Psi(x)$ associated with the asymmetric Dirac equation that will be derived in Sec. \ref{ade}. 

In other words, if we write a general homogeneous first order differential equation as
\begin{equation}
(i\hbar \gamma^\mu\partial_\mu - mc\widetilde{B})\Psi_D(x) = 0,
\end{equation}
where $\gamma^\mu$ and $\widetilde{B}$ are independent of $x^\mu$ and 
yet to be determined,
we want
\begin{equation}
(i\hbar \gamma^\mu\partial_\mu + mc\widetilde{B})
(i\hbar \gamma^\nu\partial_\nu - mc\widetilde{B})
\label{a2b2}
\end{equation}
to be equivalent to the Klein-Gordon operator
\begin{equation}
g^{\mu\nu}\partial_\mu\partial_\nu + m^2c^2/\hbar^2. 
\label{kgo}
\end{equation}
Note that the inspiration behind Dirac's approach is the simple operator relation 
$a^\mu a^\nu\partial_\mu\partial_\nu$ $-b^2 = (a^\mu\partial_\mu+b)(a^\nu\partial_\nu-b)$,
with $a^\mu$ and $b$ commuting objects independent of $x^\mu$. 
The importance of using a plus and a minus sign in the 
two factors at the right hand side is crucial to obtain a left hand side with 
no first order derivatives. 

Expanding Eq.~(\ref{a2b2}) and dropping the common factor $-\hbar^2$ we get
\begin{equation}
\frac{1}{2}\{ \gamma^\mu, \gamma^\nu\}\partial_\mu\partial_\nu 
- i(mc/\hbar)[\widetilde{B},\gamma^\mu]\partial_\mu + (m^2c^2/\hbar^2)\widetilde{B}^2,
\label{a2b2full}
\end{equation}
where $\{ X,Y \}=XY+YX$ and $[ X,Y ]=XY-YX$ are, respectively, the anticommutator and commutator of the objects $X$ and $Y$. Comparing Eq.~(\ref{a2b2full}) with (\ref{kgo}),
they are equal if
\begin{eqnarray}
\{ \gamma^\mu, \gamma^\nu\}  = 2g^{\mu\nu}, & [\widetilde{B},\gamma^\mu] = 0, 
& \widetilde{B}^2 = \mathbb{1}, 
\label{diraccondition}
\end{eqnarray}
where it is implicit that $2g^{\mu\nu}$ is multiplied by the identity operator $\mathbb{1}$
(the unity matrix in what follows).

The first condition in Eq.~(\ref{diraccondition}) is exactly the Clifford algebra defining
Dirac's gamma matrices. The second condition implies that 
$\widetilde{B}$ should be proportional
to the identity matrix $\mathbb{1}$ since this is the only class of matrices that commutes
with all gamma matrices \cite{sch05}. Finally, the third condition implies that 
$\widetilde{B} = \pm \mathbb{1}$. Both choices for the sign of 
$\widetilde{B}$ is equally valid 
and lead to a consistent description of fermions \cite{bog80} and historically Dirac chose
the plus sign, i.e., $\widetilde{B} = \mathbb{1}$.

\subsection{The asymmetric Dirac equation}
\label{ade}

Similarly to what Dirac did, we want to derive a first order homogeneous 
differential equation. However, we want this equation to give the energy-momentum relations for free particles and antiparticles associated with the Lorentz covariant Schr\"odinger 
equation \cite{rig22}. 
This is accomplished if each component of the spinor $\Psi(x)$ that solves the 
first order differential equation satisfies the Lorentz covariant Schr\"o\-din\-ger 
equation as given by Eq.~(\ref{we}). 

Contrary to the Klein-Gordon equation, Eq.~(\ref{we}) has both second and first order 
derivatives. We thus need to start from a first order differential equation and arrive at
one with first and second order derivatives. As such, Dirac's original approach,
inspired by the operator relation 
$a^\mu a^\nu\partial_\mu\partial_\nu-b^2 = (a^\mu\partial_\mu+b)(a^\nu\partial_\nu-b)$, 
has to be modified. In order to have both first and second order 
derivatives, we need both signs at the right hand side above equal, for instance, 
$(a^\mu\partial_\mu-b)(a^\nu\partial_\nu-b)$. 

As before, we write our general homogeneous first order differential equation as
\begin{equation}
(i\hbar \gamma^\mu\partial_\mu - mcB)\Psi(x) = 0.
\label{ade1}
\end{equation}
We now want 
\begin{equation}
(i\hbar \gamma^\mu\partial_\mu - mcB)
(i\hbar \gamma^\nu\partial_\nu - mcB)\Psi(x) = 0
\label{ade2}
\end{equation}
to be equivalent to  
the Lorentz covariant Schr\"odinger equation (\ref{we}),
\begin{equation}
(g^{\mu\nu}\partial_\mu\partial_\nu -i2\kappa^\mu \partial_\mu)\Psi(x)=0. 
\label{ade3}
\end{equation}
In other words, we want that each component of $\Psi(x)$ be 
a solution to the Lorentz covariant Schr\"odinger equation.

Expanding Eq.~(\ref{ade2}) and dividing by $-\hbar^2$ we get
\begin{equation}
\left(\frac{1}{2}\{ \gamma^\mu, \gamma^\nu\}\partial_\mu\partial_\nu 
+ i\widetilde{m}\{B,\gamma^\mu\}\partial_\mu - \widetilde{m}^2B^2\right)\Psi(x)=0.
\label{ade4}
\end{equation}

If we now compare Eqs.~(\ref{ade3}) and (\ref{ade4}) and demand that they should be equal,
we obtain the following three relations,
\begin{eqnarray}
\{ \gamma^\mu, \gamma^\nu\}  &=& 2g^{\mu\nu}, 
\label{c1} \\
\{B,\gamma^\mu\} &=& -2\kappa^\mu/\widetilde{m}, 
\label{c2} \\ 
B^2 &=& 0. 
\label{c3}
\end{eqnarray}

We need to find $\gamma^\mu$ and $B$ satisfying Eqs.~(\ref{c1})-(\ref{c3}) to guarantee that
the first order differential equation (\ref{ade1}) has the same dispersion relations 
of the Lorentz covariant Schr\"o\-din\-ger equation.

To solve Eq.~(\ref{c1}) we just need to assume that $\gamma^\mu$ are the usual Dirac gamma
matrices and all that remains to be done is to find the most general $B$ compatible with
Eqs.~(\ref{c2}) and (\ref{c3}).

Since the $16$ matrices $\mathbb{1},\gamma^5,\gamma^\mu,\gamma^5\gamma^\mu$, 
and $\sigma^{\mu\nu}=-\sigma^{\nu\mu}$ 
are linearly independent \cite{gre00}, an arbitrary matrix $B$
can be written as
\begin{equation}
B = b_{\mathbb{1}}\mathbb{1}  + b_5\gamma^5 + b_\mu\gamma^\mu + b_{5\mu}\gamma^5\gamma^\mu
+ \frac{b_{\mu\nu}}{2}\sigma^{\mu\nu},
\label{Bgeral}
\end{equation}
where 
\begin{eqnarray}
\gamma^5 &=& i\gamma^0\gamma^1\gamma^2\gamma^3, \label{gamma5}\\
\sigma^{\mu\nu}&=& (i/2)[\gamma^{\mu},\gamma^{\nu}] \label{smunu}, 
\end{eqnarray}
and
$b_{\mathbb{1}},b_5,b_\mu,b_{5\mu},b_{\mu\nu}$ are arbitrary complex numbers that depend
on $B$.

To compute the left hand side of Eq.~(\ref{c2}), 
we need the following anticommutators, 
\begin{eqnarray}
\{\mathbb{1}, \gamma^\mu\} &=& 2\gamma^\mu, \label{com1} \\
\{\gamma^5, \gamma^\mu\} &=& 0, \label{com2} \\
\{\gamma^\mu, \gamma^\nu\} &=& 2g^{\mu\nu}, \label{com3} \\
\{\gamma^5\gamma^\mu, \gamma^\nu\} &=& -i2\gamma^5\sigma^{\mu\nu}
=-\epsilon^{\mu\nu\alpha\beta}
g_{\eta\alpha}g_{\lambda\beta}\sigma^{\eta\lambda},\label{com4} \\
\{\sigma^{\mu\nu}, \gamma^\eta\} &=& -2\epsilon^{\eta\mu\nu\lambda}g_{\xi\lambda}
\gamma^5\gamma^\xi, \label{com5}
\end{eqnarray}
where $\epsilon^{\alpha'\beta'\mu'\nu'}=
g^{\alpha'\alpha}g^{\beta'\beta}g^{\mu'\mu}g^{\nu'\nu}\epsilon_{\alpha\beta\mu\nu}$
and $\epsilon_{\alpha\beta\mu\nu}$ is the completely antisymmetric four-dimensional 
Levi-Civita sy\-mbol, namely, $\epsilon_{\alpha\beta\mu\nu}=1$ for 
$(\alpha,\beta,\mu,\nu)=(0,1,2,3)$ and its even permutations while 
$\epsilon_{\alpha\beta\mu\nu}=-1$ for its odd permutations.

Using Eqs.~(\ref{Bgeral})-(\ref{com5}), we can write Eq.~(\ref{c2}) as
\begin{equation}
2b^\mu \!+\! 2b_{\mathbb{1}}\gamma^\mu \!-\! 2b_{\eta\lambda}g_{\alpha\nu}
\epsilon^{\mu \eta \lambda \nu} \gamma^5\gamma^\alpha 
\!+\!i2b_{5\nu}\gamma^5\sigma^{\mu\nu}
\!=\! -2\frac{\kappa^\mu}{\widetilde{m}}.
\label{c2a}
\end{equation}
The right hand side of Eq.~(\ref{c2a}) is proportional to the identity matrix.
On the other hand, in the left hand side we only have $2b^\mu$ proportional to it, 
with the other coefficients multiplying the linearly independent terms 
$\gamma^\mu,\gamma^5\gamma^\alpha$, and $\gamma^5\sigma^{\mu\nu}$.
Thus, a little algebra 
making use of the properties of $\epsilon^{\alpha\beta\mu\nu}$ necessarily implies that 
\begin{eqnarray}
b_\mu &=& -\kappa_\mu/\widetilde{m}, 
\label{bmu} \\
b_{\mathbb{1}} &=& b_{5\mu} = b_{\mu\nu} = 0.
\label{bmu2}
\end{eqnarray}

If the coefficients defining $B$ in Eq.~(\ref{Bgeral}) are given by Eqs.~(\ref{bmu})
and (\ref{bmu2}),
we have relation (\ref{c2}) satisfied and $B$ can be written as
\begin{equation}
B =  b_5\gamma^5 + b_\mu\gamma^\mu = b_5\gamma^5 -(\kappa_\mu/\widetilde{m}) \gamma^\mu.
\label{Bgeral2}
\end{equation}

What remains to be done is to check if Eq.~(\ref{Bgeral2}) also satisfies the last condition,
Eq.~(\ref{c3}). Inserting Eq.~(\ref{Bgeral2}) into (\ref{c3}) we get
\begin{equation}
(b_5)^2 + b_\mu b^\mu = 0, 
\label{c3a}
\end{equation}
where we used that $(\gamma^5)^2=\mathbb{1}$, $\{\gamma^5,\gamma^\mu\}=0$, and
Eq.~(\ref{com3}) to arrive at Eq.~(\ref{c3a}). Since Eq.~(\ref{bmu}) implies that
$b_\mu b^\mu = \kappa_\mu\kappa^\mu/\widetilde{m}^2=\kappa^2/\widetilde{m}^2$, we also have
\begin{equation}
(b_5)^2= - \kappa^2/\widetilde{m}^2.
\label{c3b}
\end{equation}

Therefore, for $B$ given by Eq.~(\ref{Bgeral2}), with $b_\mu$ and $b_5$ given by
Eqs.~(\ref{bmu}) and (\ref{c3b}), we have all the three relations given by 
Eqs.~(\ref{c1})-(\ref{c3}) satisfied.  

Moreover, using Eq.~(\ref{dispKG}) we see that Eq.~(\ref{c3b}) implies that
\begin{equation}
b_5 = \pm i. 
\end{equation}
And similarly to the choice of the sign of $\widetilde{B}$ for the standard Dirac equation \cite{bog80}, it is not difficult to see that here both choices 
for the sign of $b_5$ are equally legitimate, leading to consistent theories. For definiteness, we stick with the positive sign and from now on 
\begin{equation}
b_5 = i.
\label{b5}
\end{equation}

Using Eqs.~(\ref{bmu}), (\ref{Bgeral2}), and (\ref{b5}),  we can finally write the 
asymmetric Dirac equation as follows,
\begin{equation}
i\hbar \gamma^\mu\partial_\mu\Psi(x) - mc(i\gamma^5+b_\mu \gamma^\mu)\Psi(x) = 0.
\label{adeFinal}
\end{equation}
Or, equivalently, as
\begin{equation}
\boxed{i\hbar \gamma^\mu\partial_\mu\Psi(x) - (imc\gamma^5 - 
\hbar \kappa_\mu \gamma^\mu)\Psi(x) = 0,}
\label{adeFinal2}
\end{equation}
with $m$ given by Eq.~(\ref{restmass}). We emphasize once more that $\kappa^\mu$ 
are four relativistic invariants (scalars) of the present theory under 
proper Lorentz transformations. 
They do not transform as a four-vector and are related to the rest mass of the particle 
according to Eq.~(\ref{restmass}).

It is worth mentioning that if, instead of demanding Eq. (\ref{ade2}) to be equivalent to
the Lorentz covariant Schr\"odinger equation, we demanded that 
\begin{equation}
(i\hbar \gamma^\mu\partial_\mu + mcB)
(i\hbar \gamma^\nu\partial_\nu - mcB)\Psi(x) = 0
\label{ade2alt}
\end{equation}
should be equivalent to it, we would not succeed in obtaining a 
first order differential equation satisfying the requirements laid down
at the beginning of this section. 

Repeating the 
steps that led to Eq.~(\ref{c2}), using Eq.~(\ref{ade2alt})
instead of (\ref{ade2}), gives
\begin{equation}
[B,\gamma^\mu] = 2\kappa^\mu/\widetilde{m}. 
\label{c2alt}
\end{equation}
As we show next, Eq.~(\ref{c2alt}) cannot be satisfied for any $B$. 

To prove that, we need the following commutators,
\begin{eqnarray}
[\mathbb{1}, \gamma^\mu ] &=& 0, \label{com1a} \\
\mbox{[} \gamma^5 , \gamma^\mu ] & = & 2\gamma^5 \gamma^\mu, \label{com2a} \\
\mbox{[}\gamma^\mu, \gamma^\nu] &=& i2\sigma^{\nu\mu}, \label{com3a} \\
\mbox{[}\gamma^5\gamma^\mu, \gamma^\nu] &=& 2g^{\mu\nu}\gamma^5, \label{com4a} \\
\mbox{[}\sigma^{\mu\nu}, \gamma^\eta] &=& i2(g^{\eta\nu}\gamma^\mu 
- g^{\eta\mu}\gamma^\nu). \label{com5a}
\end{eqnarray}
Looking at Eqs.~(\ref{com1a})-(\ref{com5a}) we realize that none of them is proportional
to the identity matrix $\mathbb{1}$. This implies that we cannot satisfy Eq.~(\ref{c2alt}),
no matter how general we choose $B$ as given by Eq.~(\ref{Bgeral}). Indeed, Eqs.~(\ref{com1a})-(\ref{com5a}) tell us that the left
hand side of (\ref{c2alt}) has no term proportional to $\mathbb{1}$ while its
right hand side has a single term proportional to $\mathbb{1}$. Therefore, there is
no $B$ that satisfies (\ref{c2alt}) for massive particles since in this case we must
have at least $\kappa^0 \neq  0$.

\section{Relativistic properties}
\label{rp}

So far we have shown that it is possible to have a first order differential equation,
which we called asymmetric Dirac equation,
whose free particle-antiparticle energy-momentum relations are the ones given by
the Lorentz covariant Schr\"o\-din\-ger equation, a second order differential equation. 
But this is just half of the story. 
We now need to prove that the asymmetric Dirac equation is covariant under proper 
Lorentz transformations and calculate explicitly how the wave function
$\Psi(x)$ changes under those transformations. 

To achieve this goal, we will first obtain the conserved four-current associated with
the asymmetric Dirac equation and postulate that it should transform as a four-vector,
limiting the choices of how $\Psi(x)$ should transform under a proper Lorentz
transformation. The other constraints will appear when we require
the asymmetric Dirac equation to be covariant under those transformations.
 
\subsection{The conserved four-current}
\label{cfc}

Since $b_\mu$ is real, $\gamma^0(\gamma^5)^\dagger\gamma^0=-\gamma^5$, and 
$\gamma^0(\gamma^\mu)^\dagger\gamma^0=\gamma^\mu$, it is not difficult to see that
\begin{equation}
B^\dagger = \gamma^0 B \gamma^0.
\label{Bdagger}
\end{equation}

Using Eq.~(\ref{Bdagger}), the adjoint of the asymmetric Dirac equation (\ref{ade1})
becomes
\begin{equation}
i\hbar \partial_\mu\overline{\Psi}(x)\gamma^\mu + mc\overline{\Psi}(x)B = 0,
\label{ade1adj}
\end{equation}
where
\begin{equation}
\overline{\Psi}(x) = \Psi^\dagger(x)\gamma^0.
\end{equation}

Multiplying (\ref{ade1}) by $\overline{\Psi}(x)$ at the left, (\ref{ade1adj}) by 
$\Psi(x)$ at the right, and then summing the two expressions we get
\begin{equation}
\partial_\mu[\overline{\Psi}(x)\gamma^\mu\Psi(x)] = \partial_\mu j^\mu(x) = 0.
\label{conservedCur}
\end{equation}

Equation (\ref{conservedCur}) tells us that the asymmetric Dirac equation conserved 
four-current, $j^\mu(x) = \overline{\Psi}(x)\gamma^\mu\Psi(x)$, is formally equal to the one coming from the standard Dirac equation. 
And since we want to develop the present theory as close as possible to what we have
for the standard Dirac equation, we postulate that the four-current $j^\mu(x)$ transforms
as a contravariant four-vector after a proper Lorentz transformation. 
That is,
we assume
\begin{equation}
j^{\mu'}(x') = \Lambda^{\!\!\mu}_{\;\nu} j^\nu(x),
\label{jmu}
\end{equation}
where $\Lambda^{\!\!\mu}_{\;\nu}$ represents an arbitrary 
proper Lorentz transformation,
\begin{eqnarray}
x^{\mu'} &=& \Lambda^{\!\!\mu}_{\;\nu} x^\nu, \label{lor1} \\
\Lambda^{\;\mu}_{\!\alpha}\Lambda^{\!\!\alpha}_{\;\nu} & = & g^\mu_\nu = \delta^\mu_\nu.
\label{lor2}
\end{eqnarray}
Equation (\ref{lor2}) is a consequence of the invariance of $x_\mu x^\mu$ under a 
proper Lorentz transformation. Rewriting Eq.~(\ref{jmu}) using the definition of $j^\mu(x)$ gives
\begin{equation}
\overline{\Psi}\,'(x')\gamma^\mu\Psi'(x') = \overline{\Psi}(x)\Lambda^{\!\!\mu}_{\;\nu} 
\gamma^\nu\Psi(x).
\label{jmu2}
\end{equation}

We also write the transformed spinor $\Psi'(x')$ as
\begin{equation}
\Psi'(x') = M(x)\Psi(x),
\label{MPsi}
\end{equation}
where $M(x)$, as implied by the notation, can depend on $x$ and is assumed
to be an invertible matrix.

If we insert Eq.~(\ref{MPsi}) into the left hand side of (\ref{jmu2}) we get
\begin{equation}
\overline{\Psi}\,'(x')\gamma^\mu\Psi'(x') = \overline{\Psi}(x)\gamma^0 M^\dagger(x)
\gamma^0 \gamma^\mu M(x) \Psi(x).
\label{jmu2alt}
\end{equation}
Comparing Eqs.~(\ref{jmu2}) and (\ref{jmu2alt}) we immediately obtain
\begin{equation}
\Lambda^{\!\mu}_{\;\nu} \gamma^\nu = \gamma^0 M^\dagger(x)
\gamma^0 \gamma^\mu M(x),
\end{equation}
which, with the help of Eq.~(\ref{lor2}), becomes
\begin{equation}
\gamma^\nu = [\gamma^0 M^\dagger(x) \gamma^0] \gamma^\mu M(x) \Lambda_{\!\mu}^{\;\nu}.
\label{cond0}
\end{equation}
Equation (\ref{cond0}) is one of the constraints the matrix $M(x)$ has to satisfy.

\subsection{Relativistic covariance}

In the inertial reference frame $S'$ the asymmetric Dirac equation should look like
\begin{equation}
i\hbar \gamma^\mu\partial_{\mu'}\Psi'(x') - mcB\Psi'(x') = 0.
\label{ade1'}
\end{equation}
If we use Eq.~(\ref{MPsi}) and left multiply Eq.~(\ref{ade1'}) 
by $M^{-1}(x)$, the inverse of $M(x)$, 
we get
%
\begin{eqnarray}
&&i\hbar M^{-1}(x)\gamma^\mu M(x)\Lambda_{\!\mu}^{\;\nu}\partial_{\nu}\Psi(x) 
\nonumber \\ &&+\{i\hbar M^{-1}(x)\gamma^\mu \Lambda_{\!\mu}^{\;\nu}[\partial_{\nu}M(x)]
\nonumber \\
&&-mcM^{-1}(x)BM(x)\}\Psi(x) = 0.
\label{ade2'}
\end{eqnarray}
%

Comparing Eq.~(\ref{ade2'}) with the asymmetric Dirac equation in reference frame S,
Eq. (\ref{ade1}), they look the same (covariance) if
\begin{eqnarray}
M^{-1}\gamma^\mu M\Lambda_{\!\mu}^{\;\nu} & = & \gamma^\nu, \label{cond1} \\
i\hbar M^{-1}\gamma^\mu \Lambda_{\!\mu}^{\;\nu}\partial_{\nu}M 
-mcM^{-1}BM & = & -mcB, \label{cond2}
\end{eqnarray}
where from now on we drop the explicit reference to the dependence of $M$ on $x$.
If we left multiply Eq.~(\ref{cond2}) by $M$ we can rewrite it as
\begin{equation}
i\gamma^\mu \Lambda_{\!\mu}^{\;\nu}\partial_{\nu}M  = \widetilde{m}[B,M].
\label{cond2a}
\end{equation}

Furthermore, if we compare Eq.~(\ref{cond1}) with (\ref{cond0}), we get that the constraint
(\ref{cond0}) is equivalent to 
\begin{equation}
M^{-1} = \gamma^0 M^\dagger \gamma^0,
\label{cond3}
\end{equation}
which implies that
\begin{equation}
\overline{\Psi}\,' = \overline{\Psi}M^{-1}.
\label{psibar'}
\end{equation}

Putting everything together, the asymmetric Dirac equation is covariant under a proper
Lorentz transformation and $j^\mu$ transforms as a four-vector if there exists an 
invertible matrix $M$ such that it satisfies Eqs.~(\ref{cond1}), (\ref{cond2a}), and (\ref{cond3}). Our goal in what follows is to explicitly obtain $M$.

\subsection{Obtaining $\mathbf{M(x)}$: infinitesimal transformations}

Since we are dealing with a continuous symmetry, we will first obtain $M(x)$ for 
infinitesimal proper Lorentz transformations and only afterwards the finite ones.

Using the notation already introduced in Eqs.~(\ref{lt}) and (\ref{lor1}), an
infinitesimal proper Lorentz transformation is such that
\begin{equation}
\Lambda_{\!\mu}^{\;\nu} = g_{\mu}^{\nu} + \epsilon_{\!\mu}^{\;\nu}
\label{infLor}
\end{equation}
and the respective infinitesimal matrix $M(x)$ and its inverse are
\begin{eqnarray}
M(x) &=& \mathbb{1} + \frac{1}{2}\epsilon_{\mu\nu}M^{\mu\nu}(x), \label{infM} \\
M^{-1}(x) &=& \mathbb{1} - \frac{1}{2}\epsilon_{\mu\nu}M^{\mu\nu}(x). \label{infMinv}
\end{eqnarray}
Here $\epsilon_{\mu\nu}$ and $M^{\mu\nu}(x)$ are antisymmetric objects \cite{gre00,man86},
with the latter
being a matrix (operator) in the spinorial space.

If we insert Eqs.~(\ref{infLor})-(\ref{infMinv}) into Eq.~(\ref{cond1}), the first of the three relations $M(x)$ has to satisfy, we obtain to first order in $\epsilon_{\mu\nu}$,
\begin{equation}
[M^{\mu\nu},\gamma^\lambda] = \gamma^\mu g^{\lambda\nu} - \gamma^\nu g^{\lambda\mu}. 
\label{cfirst}
\end{equation}
Note that Eq.~(\ref{cfirst}) is the same expression we get when dealing with the original
Dirac equation. Our goal here, however, is to obtain the most general $M^{\mu\nu}$ that 
satisfies (\ref{cfirst}) before we tackle the other two relations that $M^{\mu\nu}$ 
must also satisfy.

The most general way of writing $M^{\mu\nu}(x)$ is
\begin{eqnarray}
M^{\mu\nu}(x) &=& c_{\mathbb{1}}^{\mu\nu}(x)\mathbb{1}  + c_5^{\mu\nu}(x)\gamma^5 + 
 d^{\mu\nu}_{\;\;\;\;\alpha}(x)\gamma^\alpha  \nonumber \\
&& + f^{\mu\nu}_{\;\;\;\;\alpha}(x)\gamma^5\gamma^\alpha+ \frac{h^{\mu\nu}_{\;\;\;\;\alpha\beta}(x)}{2}\sigma^{\alpha\beta},
\label{MmnGeral}
\end{eqnarray}
where, due to the fact that $\sigma^{\alpha\beta}$ is antisymmetric, 
we can work without losing in generality with $h^{\mu\nu}_{\;\;\;\;\alpha\beta}$ 
antisymmetric in the indexes $\alpha,\beta$, i.e., 
$h^{\mu\nu}_{\;\;\;\;\alpha\beta} = -h^{\mu\nu}_{\;\;\;\;\beta\alpha}$.
In addition to that, since $M^{\mu\nu}$ is also antisymmetric, all the coefficients 
appearing in Eq.~(\ref{MmnGeral}) are antisymmetric in the indexes $\mu,\nu$.

Inserting Eq.~(\ref{MmnGeral}) into the left hand side of (\ref{cfirst}) and using the
commutators listed in Eqs.~(\ref{com1a})-(\ref{com5a}) we get
%
\begin{eqnarray}
[2f^{\mu\nu\lambda}(x)]\gamma^5 + [2c_5^{\mu\nu}(x)]\gamma^5\gamma^\lambda
+ [i2d^{\mu\nu}_{\;\;\;\;\alpha}(x)]\sigma^{\lambda\alpha}\nonumber \\
+ ih^{\mu\nu}_{\;\;\;\;\alpha\beta}(x)
(\gamma^\alpha g^{\lambda\beta} - \gamma^\beta g^{\lambda\alpha}) 
= \gamma^\mu g^{\lambda\nu} - \gamma^\nu g^{\lambda\mu}.
\label{cfirst2}
\end{eqnarray}
%
%
Since $\gamma^5,\gamma^5\gamma^\mu,\sigma^{\mu\nu}$ and $\gamma^\mu$ are linear independent,
it is not difficult to see that Eq.~(\ref{cfirst2}) implies that
\begin{equation}
c_5^{\mu\nu}(x) = d^{\mu\nu}_{\;\;\;\;\alpha}(x) = f^{\mu\nu}_{\;\;\;\;\alpha}(x) = 0
\label{cfirst3}
\end{equation}
and
\begin{equation}
h^{\mu\nu}_{\;\;\;\;\alpha\beta}(x) = -\frac{i}{2}(\delta^\mu_\alpha\delta^\nu_\beta
-\delta^\mu_\beta\delta^\nu_\alpha).
\label{cfirst4}
\end{equation}
If we now insert Eqs.~(\ref{cfirst3}) and (\ref{cfirst4}) into (\ref{MmnGeral}) 
we obtain
\begin{equation}
M^{\mu\nu}(x) = c_{\mathbb{1}}^{\mu\nu}(x)\mathbb{1}  
- \frac{i}{2}\sigma^{\mu\nu}.
\label{MmnGeral3}
\end{equation}

Moving to the second relation that $M(x)$ has to satisfy, we obtain to first order in
$\epsilon_{\mu\nu}$,  
\begin{equation}
i\gamma^\mu \partial_{\mu}M^{\alpha\beta}  = \widetilde{m}[B,M^{\alpha\beta}],
\label{csecond}
\end{equation}
after we insert Eqs.~(\ref{infLor})-(\ref{infM}) into (\ref{cond2a}).
Using Eqs.~(\ref{Bgeral2}), (\ref{com5a}), (\ref{MmnGeral3}), and that
$[\gamma^5,\sigma^{\mu\nu}]=0$, we can write Eq.~(\ref{csecond}) as
\begin{equation}
\partial_{\mu} c_{\mathbb{1}}^{\alpha\beta}(x) = i\widetilde{m}
(b^\beta\delta^\alpha_\mu - b^\alpha\delta^\beta_\mu),
\end{equation}
whose solution is  
\begin{equation}
c_{\mathbb{1}}^{\alpha\beta}(x) = i\widetilde{m} (b^\nu x^\mu - b^\mu x^\nu) 
+ c^{\mu\nu}_o
\label{csecond2}
\end{equation}
if we remember that $\partial_\mu x^\nu = \delta^\nu_\mu$. Here $c^{\mu\nu}_o$
is an arbitrary constant.

Using Eqs.~(\ref{bmu}) and (\ref{csecond2}), Eq.~(\ref{MmnGeral3}) becomes
\begin{equation}
M^{\mu\nu}(x) = c^{\mu\nu}_o + i (\kappa^\mu x^\nu - \kappa^\nu x^\mu)  
- \frac{i}{2}\sigma^{\mu\nu}.
\label{MmnGeral4}
\end{equation}

The last constraint on $M(x)$, Eq.~(\ref{cond3}), can be written as 
\begin{equation}
M^{\mu\nu} = -\gamma^0 (M^{\mu\nu})^\dagger \gamma^0
\label{cthird}
\end{equation}
after Eqs.~(\ref{infM}) and (\ref{infMinv}). Using Eq.~(\ref{MmnGeral4}),
noting that 
$(\sigma^{\mu\nu})^\dagger=\gamma^0\sigma^{\mu\nu}\gamma^0$,
and remembering that $\kappa^\mu$ is real, Eq.~(\ref{cthird}) gives
\begin{equation}
c^{\mu\nu}_o = - (c^{\mu\nu}_o)^*.
\label{cthird2}
\end{equation}
In other words, $c^{\mu\nu}_o$ is a pure imaginary and we thus write it as
\begin{equation}
c^{\mu\nu}_o = i a^{\mu\nu},
\label{amunu}
\end{equation}
where $a^{\mu\nu}$ is a real constant.

Using Eq.~(\ref{amunu}) we can write Eq.~(\ref{MmnGeral4}) as
\begin{equation}
M^{\mu\nu}(x) = i a^{\mu\nu} + i (\kappa^\mu x^\nu - \kappa^\nu x^\mu)  
- \frac{i}{2}\sigma^{\mu\nu}
\label{MmnGeral5}
\end{equation}
and, finally, Eq.~(\ref{infM}) for an arbitrary infinitesimal proper 
Lorentz transformation,
\begin{equation}
M(x) = \mathbb{1} +  \frac{i}{2}\epsilon_{\mu\nu} a^{\mu\nu} 
+ \frac{i}{2}\epsilon_{\mu\nu} K^{\mu\nu}(x)  
- \frac{i}{4}\epsilon_{\mu\nu}\sigma^{\mu\nu},
\label{Minf}
\end{equation}
where
\begin{equation}
K^{\mu\nu}(x) = \kappa^\mu x^\nu - \kappa^\nu x^\mu.
\label{Kmunu}
\end{equation}
Note that if  $a^{\mu\nu}=\kappa^\mu=0$, 
Eq.~(\ref{Minf}) reduces to the transformation
associated with the standard Dirac equation.

\subsection{Obtaining $\mathbf{M(x)}$: finite transformations}

\subsubsection{Finite boosts}

For an infinitesimal boost along the $x^1$-axis we have 
$\epsilon^{\mu}_{\;\;\nu} = \epsilon (I_x)^{\mu}_{\;\;\nu}$, where 
$\epsilon > 0$ and the 
generator is  $(I_x)^{0}_{\;\;1}=(I_x)^{1}_{\;\;0}=-1$ and zero otherwise \cite{gre00}.
Using this notation, Eq.~(\ref{Minf}) becomes
\begin{eqnarray}
M(x) &=& \mathbb{1} +  \frac{i}{2}\epsilon (I_x)_{\mu\nu} a^{\mu\nu} 
+ \frac{i}{2}\epsilon (I_x)_{\mu\nu} K^{\mu\nu}(x) \nonumber \\ 
&&- \frac{i}{4}\epsilon (I_x)_{\mu\nu}\sigma^{\mu\nu}.
\label{MinfBoost}
\end{eqnarray}

To order
$\epsilon$ we can write Eq.~(\ref{MinfBoost}) as
\begin{eqnarray}
M(x) &=& \left[\mathbb{1} +  \frac{i}{2}\epsilon (I_x)_{\mu\nu} a^{\mu\nu}\right] 
\left[\mathbb{1} + \frac{i}{2}\epsilon (I_x)_{\mu\nu} K^{\mu\nu}(x)\right] 
\nonumber \\ 
&&\times \left[\mathbb{1}- \frac{i}{4}\epsilon (I_x)_{\mu\nu}\sigma^{\mu\nu}\right] + 
\mathcal{O}(\epsilon).
\label{MinfBoost2}
\end{eqnarray}

To go from the inertial reference frame $S$ to $S'$ moving with speed $v$ in the 
$x^1$ direction, we apply $N$ infinitesimal boosts
with rapidity $\epsilon$ \cite{gre00}, where the finite rapidity (hyperbolic angle) is
$\omega  = N \epsilon$. Denoting the reference frame $S$ and its variables using the subscript $0$, frame $S'$ and its variables using the subscript $N$, and calling
the intermediate reference frames $S_n$, we have after $N$ hyperbolic
rotations
\begin{equation}
\Psi_N(x_N) = f^N\left(\frac{\omega}{N}\right)\Pi_{j=0}^{N-1}g\left(\frac{\omega}{N}, x_j\right)h^N\left(\frac{\omega}{N}\right)\Psi_0(x_0), 
\label{MinfBoost3}
\end{equation}
where
\begin{eqnarray}
f(\epsilon) &=& \mathbb{1} +  \frac{i}{2}\epsilon (I_x)_{\mu\nu} a^{\mu\nu},
\label{f}\\
g(\epsilon,x) &=& \mathbb{1} +  \frac{i}{2}\epsilon (I_x)_{\mu\nu} K^{\mu\nu}(x),
\label{g}\\
h(\epsilon) &=& \mathbb{1} -  \frac{i}{4}\epsilon (I_x)_{\mu\nu} \sigma^{\mu\nu}.
\label{h}
\end{eqnarray}
We have also used that $a^{\mu\nu}, K^{\mu\nu}$, and $\sigma^{\mu\nu}$ commute among each other to obtain Eq.~(\ref{MinfBoost3}).

In the limit where $N\rightarrow \infty$ we obtain
\begin{eqnarray}
\Psi'(x') &=& \lim_{N\rightarrow \infty}\left[f^N\left(\frac{\omega}{N}\right)\right]
\lim_{N\rightarrow \infty}\left[\Pi_{j=0}^{N-1}g\left(\frac{\omega}{N}, x_j\right)\right]
\nonumber \\
&& \times \lim_{N\rightarrow \infty}\left[h^N\left(\frac{\omega}{N}\right)\right]\Psi_0(x_0). 
\label{MinfBoost4}
\end{eqnarray}

However, it is not difficult to see that using Eqs.~(\ref{f}) and (\ref{h}) we have \cite{gre00},
\begin{eqnarray}
\lim_{N\rightarrow \infty}\left[f^N\left(\frac{\omega}{N}\right)\right] &=& 
\exp\{(i/2)\omega a_{\mu\nu}(I_x)^{\mu\nu}\}, \label{f2}\\
\lim_{N\rightarrow \infty}\left[h^N\left(\frac{\omega}{N}\right)\right] &=& 
\exp\{-(i/4)\omega \sigma_{\mu\nu}(I_x)^{\mu\nu}\} \nonumber \\
&=& \exp\{-(i/2)\omega \sigma^{10}\}. \label{h2}
\end{eqnarray}
Equation (\ref{f2}) is related to a global constant phase, $e^{i\theta}$, 
since $a_{\mu\nu}$ is proportional to the identity matrix that acts on the spinorial space. 
We can thus  drop
this term from now on without losing in generality. Equation (\ref{h2}) is exactly the transformation law for the Dirac spinor after a boost in the $x^1$ 
direction \cite{gre00}. Henceforth we 
call it by its usual notation, i.e., $S$ (do not confuse with reference frame $S$). 

The mathematical steps needed to compute the remaining limit in Eq.~(\ref{MinfBoost4}) is a little more 
involved. This comes about since at each step we need to change the variables to the
new reference frame. Contrary to Eqs.~(\ref{f}) and (\ref{h}), Eq.~(\ref{g}) depends
on the space-time coordinate $x^\mu$. The details of this long calculation can be found in 
the appendix $C$ of ref. \cite{rig22}. The final result is
%
\begin{eqnarray}
&&K(x) = \lim_{N\rightarrow \infty}\left[\Pi_{j=0}^{N-1}g\left(\frac{\omega}{N}, x_j\right)\right] \nonumber \\
&& =  \exp \{ i[(\gamma -1)\kappa^0\!+\!\gamma \beta \kappa^1]x^0\!\!-\!
i[\gamma\beta\kappa^0
 \!+\!(\gamma -1)\kappa^1]x^1\}. \nonumber \\
&& \label{Kx}
\end{eqnarray}
%
%
where $\gamma=1/\sqrt{1-\beta^2}$ is the Lorentz factor, $\beta=v/c$, and the rapidity
$\omega$ is connected to $\beta$ by the following relation, $\tanh \omega = \beta$.
Note that in Ref. \cite{rig22} the calculation is done from reference frame $S'$ to
$S$ while here we are going from $S$ to $S'$. For boosts, this means that we should 
change $\beta$ to $-\beta$ in the expression in Ref. \cite{rig22}
to obtain Eq.~(\ref{Kx}) and for spatial rotations we should
change the sign of the rotation angle to get the results presented next.

\subsubsection{Finite spatial rotations}

The same steps adopted to obtain the finite boost transformation from $N$ infinitesimal
ones apply for finite spatial rotations too. The main difference is that instead of hyperbolic angles we now have euclidean rotation angles about a given spatial axis.

Similarly to finite boosts, we have
\begin{equation}
\Psi'(x') = M(x) \Psi(x) = K(x)S\Psi(x),
\label{Mfinite}
\end{equation}
where $S$ again is the corresponding transformation law for a Dirac spinor subjected
to a given proper Lorentz transformation, a spatial rotation in the present scenario, 
and $K(x)$ is the
corresponding transformation law of the Lorentz-Schr\"o\-din\-ger  
scalar wave function $\psi(x)$, 
where $\psi(x)$ satisfies the Lorentz covariant Schr\"o\-din\-ger
equation \cite{rig22}.

We illustrate the previous point showing the explicit finite transformation law for 
two cases, namely, rotations about two different orthogonal axes, which are enough to 
generate any finite rotation. 

For a $\varphi$ radian 
rotation about the $x^3$-axis ($z$-axis), whose generator is 
$(I_3)^{1}_{\;\;2} =-(I_3)^{2}_{\;\;1}$ $=1$ and zero otherwise, we get \cite{rig22,gre00}
\begin{eqnarray}
S & = & \exp\{(i/2)\varphi \sigma^{12}\}, \\
K(x)  & = & \exp\{i[\kappa^1x^1+\kappa^2x^2](1-\cos\varphi)\} \nonumber \\
& & \times \exp\{-i[\kappa^1x^2-\kappa^2x^1]\sin\varphi\}.
\end{eqnarray}

For a
rotation of  $\varphi$ radians about the $x^1$-axis ($x$-axis), whose generator is 
$(I_1)^{2}_{\;\;3}=-(I_1)^{3}_{\;\;2}=1$ and zero otherwise, we get \cite{rig22,gre00}
\begin{eqnarray}
S & = & \exp\{(i/2)\varphi \sigma^{23}\}, \\
K(x)  & = & \exp\{i[\kappa^2x^2+\kappa^3x^3](1-\cos\varphi)\} \nonumber \\
& & \times \exp\{-i[\kappa^2x^3-\kappa^3x^2]\sin\varphi\}.
\end{eqnarray}

We have also implemented a consistency check of the validity of 
the previous calculations that led 
to the above three finite proper Lorentz transformations (one boost and two rotations).
We have checked that they all satisfy Eqs. (\ref{cond1}), (\ref{cond2a}), and
(\ref{cond3}), the three relations that any transformation $M(x)$, 
finite or infinitesimal, should satisfy.  

\textit{Remark}. Since $\Psi'(x') = K(x)S\Psi(x)$, with $S$ being exactly the 
transformation law we get when dealing with a Dirac spinor, and $K(x)$ being
proportional to the identity matrix acting on
the spinorial space, with $K^\dagger(x)=K^{-1}(x)$, the bilinears of the present 
theory transform exactly in the same way as the Dirac bilinears do after a proper Lorentz 
transformation. For improper Lorentz transformations, however,
the analysis is more subtle and
we will deal with it when we investigate the discrete symmetries of the 
asymmetric Dirac equation at the end of this work.

\section{Connection to the Dirac equation}
\label{cde}

If we insert the following ansatz,
\begin{equation}
\Psi(x) = e^{i\kappa_\mu x^\mu}U\Psi_D(x),
\label{aDtoD}
\end{equation}
where
\begin{equation}
U = \left(\frac{\mathbb{1}-i\gamma^5}{\sqrt{2}}
\right), 
\end{equation}
into the asymmetric Dirac equation (\ref{adeFinal2}) we get
\begin{equation}
e^{i\kappa_\mu x^\mu}U^\dagger
[i\hbar \gamma^\mu\partial_\mu\Psi_D(x) -mc\Psi_D(x)] = 0.
\label{dir1}
\end{equation}
To obtain Eq.~(\ref{dir1}) we used that $\gamma^\mu U = U^\dagger \gamma^\mu$, 
which follows from $\gamma^\mu \gamma^5=-\gamma^5\gamma^\mu$ and 
$\gamma^5=\gamma^{5\dagger}$, and that $\gamma^5U=-iU^\dagger$, a consequence of
$(\gamma^5)^2=1$.

If we left multiply Eq.~(\ref{dir1}) by $e^{-i\kappa_\mu x^\mu}U$ we obtain
\begin{equation}
 i\hbar \gamma^\mu\partial_\mu\Psi_D(x) -mc\Psi_D(x) = 0,
\label{dir2}
\end{equation}
which is exactly the Dirac equation. As such, $\Psi_D(x)$ is a solution to the
standard Dirac equation and Eq.~(\ref{aDtoD}) is the connection 
between the asymmetric Dirac equation and the standard one. 

As anticipated in the remark at the end of the last section, we can now better 
appreciate that the asymmetric Dirac equation's behavior under discrete symmetry
operations 
is not straightforward extensions of what we know from the Dirac equation. The 
reason for this different behavior is the presence of the $\gamma^5$ matrix in the 
unitary matrix connecting both equations [cf. Eq.~(\ref{aDtoD})]. 
Under a Lorentz transformation $\Lambda$, it is known that 
$S^{-1}\gamma^5S = \gamma^5 det\Lambda$, where $det\Lambda$ is the determinant of the 
matrix $\Lambda$ \cite{man86}. Therefore, for improper Lorentz
transformations a minus sign will show up and thus $\gamma^5 \rightarrow -\gamma^5$, 
which requires a careful analysis when we build the parity and time reversal operators 
associated with the asymmetric Dirac equation. We will come back to this
issue in Sec. \ref{ds}  and in Sec. \ref{mcp} we will further explore under what 
conditions the asymmetric Dirac equation and the Dirac equation lead to the same experimental
predictions at the first quantization level.

\section{Plane wave solutions}
\label{pws}

Inserting the ansatz
\begin{equation}
\Psi(x) = e^{i\kappa_\mu x^\mu}\widetilde{\Psi}(x)
\label{aDtoD2}
\end{equation}
into the asymmetric Dirac equation (\ref{adeFinal2}), we obtain
after multiplying by $e^{-i\kappa_\mu x^\mu}$,
\begin{equation}
i\hbar \gamma^\mu\partial_\mu\widetilde{\Psi}(x) -imc\gamma^5\widetilde{\Psi}(x) = 0.
\label{dirA}
\end{equation}

Two interesting features of Eq.~(\ref{dirA}) are the following. First, a direct calculation shows that 
\begin{equation}
(i\hbar \gamma^\mu\partial_\mu -imc\gamma^5)(i\hbar \gamma^\mu\partial_\mu -imc\gamma^5)\widetilde{\Psi}(x) = 0
\end{equation}
is
\begin{equation}
(\partial_\mu\partial^\mu + \widetilde{m}^2)\widetilde{\Psi}(x)=0,
\end{equation}
i.e., each component of Eq.~(\ref{dirA}) satisfies the Klein-Gordon equation. As such,
the dispersion relation for the plane wave solutions of Eq.~(\ref{dirA}) are the 
standard relativistic energy-momentum relation.

Second, as we show in the appendix 
\ref{apA}, 
the wave function $\widetilde{\Psi}(x)$ that solves (\ref{dirA}) transforms under
a proper Lorentz transformation in exactly the same way as the standard Dirac spinor,
\begin{equation}
\widetilde{\Psi}'(x')=S\widetilde{\Psi}(x). 
\label{PsiTildeTransf}
\end{equation}
This implies that Eq.~(\ref{dirA}) is Lorentz 
covariant, which can be seen by noting that
$S$ commutes with $\gamma^5$ (and this is true because $[\gamma^5,\sigma^{\mu\nu}]=0$).

Since the free particle solutions to Eq.~(\ref{dirA}) lead to the standard relativistic
dispersion relations, it is not difficult to see that the following ansatz is the most
convenient one to represent the positive energy plane wave solutions  of 
the asymmetric Dirac equation,
\begin{equation}
\Psi(x) = cte \; u_r(\mathbf{p})e^{i\kappa_\mu x^\mu} e^{-ip_\mu x^\mu/\hbar}. 
\label{ur}
\end{equation}
Here $cte$ is a normalization constant and $u_r(\mathbf{p})$ is a spinor, 
with $r=1,2$ labeling
two linearly independent solutions that we choose to be orthogonal.
These solutions are associated with particles with positive 
energies given by $E_\mathbf{p}^+=-\hbar c \kappa^0 + E_\mathbf{p}$, with
$E_\mathbf{p} = \sqrt{m^2c^4 + |\mathbf{p}|^2c^2}$. 

As we show next, Eq.~(\ref{ur}) when inserted into the asymmetric Dirac equation (\ref{adeFinal2}) gives a matrix equation for $u_r(\mathbf{p})$ that, as the 
notation implies, does not depend on $\kappa^\mu$. Also, $k^\mu = p^\mu/\hbar$
satisfies the same relations of the Dirac's relativistic four-wave vector 
and are consistent with the 
usual interpretations associated with it.

Inserting Eq.~(\ref{ur}) into (\ref{adeFinal2}) leads to
\begin{equation}
(\slashed p - i m c \gamma^5)u_r(\mathbf{p}) = 0
\label{ue}
\end{equation}
and, consequently, to its adjoint equation,
\begin{equation}
\overline{u}_r(\mathbf{p})(\slashed p - i m c \gamma^5) = 0, 
\label{ubare}
\end{equation}
where
\begin{equation}
\overline{u}_r(\mathbf{p}) = u^\dagger_r(\mathbf{p})\gamma^0 
\label{uadj}
\end{equation}
and $\slashed p = \gamma^\mu p_\mu$ is Feynman slash notation.

A direct calculation using that $\slashed p \slashed p = p_\mu p^\mu = p^2$ and
$\gamma^5\gamma^\mu = - \gamma^\mu\gamma^5$ gives 
\begin{equation}
(\slashed p - i m c \gamma^5)^2 u_r(\mathbf{p}) = (p^2 - m^2 c^2) u_r(\mathbf{p})= 0.
\label{usquared}
\end{equation}
Equation (\ref{usquared}) implies that $p^2=m^2c^2$, i.e, the standard relativistic
energy-momentum relation. Being more explicit, we have $(p^0)^2=|\mathbf{p}|^2+m^2c^2$,
with $p^0=E_{\mathbf{p}}/c$. Note, however, that the energy and momentum for particles of mass $m$ described by the asymmetric Dirac equation are, respectively, 
$E_\mathbf{p}^+=-\hbar c \kappa^0 + E_\mathbf{p}$ and $p_+^j=-\hbar\kappa^j + p^j$, with $p^j=\hbar k^j$.

To describe the plane wave solutions with ``negative'' energies, i.e., antiparticles with positive energies when we second quantize the asymmetric Dirac equation \cite{rig23}, the 
following ansatz is the most convenient one,
\begin{equation}
\Psi(x) = cte \; v_r(\mathbf{p})e^{i\kappa_\mu x^\mu} e^{ip_\mu x^\mu/\hbar}. 
\label{vr}
\end{equation}
As before, $cte$ is a normalization constant and $v_r(\mathbf{p})$ is a spinor 
with $r=1,2$ labeling two linearly independent solutions that we choose to be orthogonal.
Now, however, these solutions are related to antiparticles with positive 
energies given by $E_\mathbf{p}^-=\hbar c \kappa^0 + E_\mathbf{p}$.

Similarly to the calculations involving $u_r(\mathbf{p})$, 
it is not difficult to see that
we now have
\begin{eqnarray}
(\slashed p + i m c \gamma^5)v_r(\mathbf{p}) &=& 0, \label{ve} \\
\overline{v}_r(\mathbf{p})(\slashed p + i m c \gamma^5) &=& 0, 
\label{vbare}
\end{eqnarray}
where
\begin{equation}
\overline{v}_r(\mathbf{p}) = v^\dagger_r(\mathbf{p})\gamma^0. 
\label{vadj}
\end{equation}
Also,
\begin{equation}
(\slashed p + i m c \gamma^5)^2 v_r(\mathbf{p}) = (p^2 - m^2 c^2) v_r(\mathbf{p})= 0,
\label{vsquared}
\end{equation}
which implies that $p^2=m^2c^2$ and $(p^0)^2=|\mathbf{p}|^2+m^2c^2$. 
The energy and momentum for antiparticles of mass $m$ described by the asymmetric Dirac equation are, respectively, 
$E_\mathbf{p}^-=\hbar c \kappa^0 + E_\mathbf{p}$ and $p_-^j=\hbar\kappa^j + p^j$, 
with $p^j=\hbar k^j$.

We should mention that which vacuum excitation we call particle or antiparticle is 
rather arbitrary. Following the choice adopted in Ref. \cite{rig22}, we call particles
the vacuum excitations with the smallest energy for given a wave number $k^j$ and 
antiparticles the excitations with the greatest energy for the same wave number. 

Following the standard prescription \cite{gre00,man86}, 
we normalize $u_r(\mathbf{p})$ and 
$v_r(\mathbf{p})$ as follows,
\begin{equation}
u_r^\dagger(\mathbf{p})u_r(\mathbf{p}) = v_r^\dagger(\mathbf{p})v_r(\mathbf{p}) =
\frac{E_{\mathbf{p}}}{mc^2}.
\end{equation}
Since for a degenerate eigenvalue we can always choose orthogonal eigenvectors, we have
\begin{equation}
u_r^\dagger(\mathbf{p})u_s(\mathbf{p}) = v_r^\dagger(\mathbf{p})v_s(\mathbf{p}) =
\frac{E_{\mathbf{p}}}{mc^2}\delta_{rs}=\frac{p^0}{mc}\delta_{rs},
\label{urvs}
\end{equation}
where $\delta_{rs}$ is the Kronecker delta.

As we prove in the appendix 
\ref{apB}, 
Eqs. (\ref{ue}), (\ref{ubare}), (\ref{ve}), (\ref{vbare}), and
(\ref{urvs}) lead to the following orthonormality relations, 
\begin{eqnarray}
u_r^\dagger(\mathbf{p})v_s(\mathbf{-p}) =& v_r^\dagger(\mathbf{p})u_s(\mathbf{-p}) 
&=  0, \label{o1}\\
\overline{u}_r(\mathbf{p})u_s(\mathbf{p}) =& 
\overline{v}_r(\mathbf{p})v_s(\mathbf{p}) 
&= 0, \label{o2}\\
\overline{u}_r(\mathbf{p})\gamma^5v_s(\mathbf{p}) =& 
\overline{v}_r(\mathbf{p})\gamma^5u_s(\mathbf{p}) 
&=  0, \label{o3}\\
\overline{u}_r(\mathbf{p})i\gamma^5u_s(\mathbf{p}) =& 
-\overline{v}_r(\mathbf{p})i\gamma^5v_s(\mathbf{p}) 
&=  \delta_{rs}, \label{o4}
\end{eqnarray}
to the following completeness relation (resolution of the identity),
\begin{eqnarray}
\mathbb{1} &=& \sum_{r=1}^2[u_r(\mathbf{p})\overline{u}_r(\mathbf{p})i\gamma^5
 - v_r(\mathbf{p})\overline{v}_r(\mathbf{p})i\gamma^5] \label{complete1}\\
 &=& \sum_{r=1}^2[i\gamma^5u_r(\mathbf{p})\overline{u}_r(\mathbf{p})
 - i\gamma^5v_r(\mathbf{p})\overline{v}_r(\mathbf{p})], \label{complete2}
\end{eqnarray}
and to these identities,
\begin{eqnarray}
\overline{u}_s(\mathbf{-p})v_r(\mathbf{p}) &=& -\frac{ip^0}{mc}
u_s^\dagger(\mathbf{-p})\gamma^5 v_r(\mathbf{p}), \label{i1}\\
\overline{v}_s(\mathbf{-p})u_r(\mathbf{p}) &=& \frac{ip^0}{mc}
v_s^\dagger(\mathbf{-p})\gamma^5 u_r(\mathbf{p}). \label{i2}
\end{eqnarray}

It is worth noting that several of the previous relations are different from the 
ones the Dirac spinors satisfy. The standard Dirac spinors 
$u_{\!_Dr}(\mathbf{p})$ and $v_{\!_Dr}(\mathbf{p})$ do not satisfy 
Eqs. (\ref{o2})-(\ref{i2}) for all $r$ and $s$. 
For instance, they do not satisfy Eq.~(\ref{o2}) for $r=s$ and 
the equivalent expressions related to Eqs.~(\ref{o3}) and (\ref{o4}) do not have the 
$\gamma^5$ matrix and the imaginary number $i$ \cite{man86}.

\section{Energy projection operators}

The energy projection operators for the asymmetric Dirac equation are
\begin{equation}
\Lambda^{\pm}(\mathbf{p}) = \frac{\gamma^5(\pm \slashed p + imc\gamma^5)}{2imc}
= \frac{(\pm \slashed p - imc\gamma^5)i\gamma^5}{2mc},
\label{epo}
\end{equation}
where we used that $\gamma^5\slashed p =-\slashed p \gamma^5$ to obtain the last term
above from the middle one.

Using Eqs.~(\ref{ue})-(\ref{ubare}) and  (\ref{ve})-(\ref{vbare}) we can prove that
\begin{eqnarray}
\Lambda^+(\mathbf{p}) u_r(\mathbf{p}) &=& u_r(\mathbf{p}), \label{l+u}\\
\Lambda^+(\mathbf{p}) v_r(\mathbf{p}) &=& 0, \label{l+v}\\
\Lambda^-(\mathbf{p}) v_r(\mathbf{p}) &=& v_r(\mathbf{p}), \label{l-v} \\
\Lambda^-(\mathbf{p}) u_r(\mathbf{p}) &=& 0. \label{l-u}
\end{eqnarray}
Equations (\ref{l+u})-(\ref{l-u}) are equal to the relations obtained for the standard
Dirac equation. 

To prove Eq.~(\ref{l+u}) we add and subtract $imc\gamma^5$ inside the parenthesis that appear in the definition of $\Lambda^+$ [cf. Eq.~(\ref{epo})]. Then, we proceed as follows,
\begin{eqnarray}
\Lambda^+(\mathbf{p}) u_r(\mathbf{p}) &=& 
\frac{\gamma^5(\slashed p - imc\gamma^5+ 2imc\gamma^5)}{2imc}u_r(\mathbf{p}) 
\nonumber \\
&=& \frac{\gamma^5}{2imc}(\slashed p - imc\gamma^5)u_r(\mathbf{p}) + (\gamma^5)^2 u_r(\mathbf{p}) \nonumber \\
&=& 0 + \mathbb{1}u_r(\mathbf{p}) = u_r(\mathbf{p}),
\end{eqnarray}
where Eq.~(\ref{ue}) was used to obtain the last line. To see that Eq.~(\ref{l+v})
is indeed true, we just need to use Eq.~(\ref{ve}). In a similar way we prove 
Eqs.~(\ref{l-v}) and (\ref{l-u}).

On the other hand, the action of $\Lambda^\pm(\mathbf{p})$ on the adjoint spinors 
are not the same
we obtain for the standard Dirac equation. Here we have
\begin{eqnarray}
\overline{u}_r(\mathbf{p}) \Lambda^+(\mathbf{p})  &=& 0, \label{l+ubar}\\
\overline{v}_r(\mathbf{p}) \Lambda^+(\mathbf{p}) &=& \overline{v}_r(\mathbf{p}), 
\label{l+vbar}\\
\overline{v}_r(\mathbf{p})\Lambda^-(\mathbf{p}) &=& 0, \label{l-vbar} \\
\overline{u}_r(\mathbf{p})\Lambda^-(\mathbf{p}) &=& \overline{u}_r(\mathbf{p}). 
\label{l-ubar}
\end{eqnarray}
The same techniques used to prove Eqs.~(\ref{l+u})-(\ref{l-u}) apply here. We just need 
to use Eqs.~(\ref{ubare}) and (\ref{vbare}) instead of (\ref{ue}) and (\ref{ve}) 
to complete the proofs. The corresponding expressions for the Dirac spinors are
\cite{man86}, 
$\overline{u}_{\!_Dr} \Lambda^+_D  =\overline{u}_{\!_Dr}$,
$\overline{v}_{\!_Dr} \Lambda^+_D  =0,
\overline{v}_{\!_Dr} \Lambda^-_D  =\overline{v}_{\!_Dr}
$, and
$
\overline{u}_{\!_Dr} \Lambda^-_D  =0,
$
where
$
\Lambda^\pm_D = (\pm \slashed p + mc)/(2mc).
$

Note that if we look at Eq.~(\ref{o2}), 
we can also understand why we must get Eqs.~(\ref{l+ubar}) and (\ref{l-vbar}). 
For instance, Eq.~(\ref{o2}) tells us that 
$\overline{u}_r(\mathbf{p})$ is orthogonal to $u_s(\mathbf{p})$, for any $r,s$.
Therefore, $u_s(\mathbf{p})$ and $\overline{u}_r(\mathbf{p})$ must be associated with
states having different eigenenergies. As such, if
$\Lambda^+(\mathbf{p})u_r(\mathbf{p})\neq 0$ 
it is expected that $\overline{u}_r(\mathbf{p})\Lambda^+(\mathbf{p})=0$.
A similar reasoning can be made to explain Eq.~(\ref{l-vbar}).

Using that $\slashed p \slashed p = p^2=m^2c^2$,  
$\slashed p \gamma^5 = -\gamma^5 \slashed p$, and $(\gamma^5)^2=\mathbb{1}$, 
we can also show that $\Lambda^\pm(\mathbf{p})$
satisfies the usual properties of projection operators,
\begin{eqnarray}
[\Lambda^\pm(\mathbf{p})]^2 & = & \Lambda^\pm(\mathbf{p}), \\
\Lambda^\pm(\mathbf{p})\Lambda^\mp(\mathbf{p})&=& 0 \\
\Lambda^+(\mathbf{p}) + \Lambda^-(\mathbf{p})&=& \mathbb{1}.
\end{eqnarray}

Finally, if we apply the completeness relation (\ref{complete1}) at the left and 
right of $\Lambda^\pm(\mathbf{p})$ we get
\begin{eqnarray}
\Lambda^+(\mathbf{p}) & = & \sum_{r=1}^2u_r(\mathbf{p})\overline{u}_r(\mathbf{p})i\gamma^5,
\label{complete+} \\
\Lambda^-(\mathbf{p}) & = & -\sum_{r=1}^2v_r(\mathbf{p})\overline{v}_r(\mathbf{p})i\gamma^5.
\label{complete-}
\end{eqnarray}

\section{Helicity and spin projection operators}

\subsection{Helicity projection operators}

Similarly to the standard Dirac equation, \textit{helicity} 
is another quantum number that can be used to classify the free particle solutions 
of the asymmetric Dirac equation.
We define it  here in exactly the same way we do for the standard
Dirac equation. The helicity projection operators are \cite{gre00,man86},
\begin{equation}
\Pi^\pm(\mathbf{p}) = \frac{1}{2}(\mathbb{1} \pm \sigma_{\mathbf{p}}),
\label{hpo}
\end{equation}
where
\begin{equation}
\sigma_{\mathbf{p}} = \frac{\bm{\sigma}\cdot \mathbf{p}}{|\mathbf{p}|}
\label{sigmap}
\end{equation}
and
\begin{equation}
\bm{\sigma} = (\sigma^{23},\sigma^{31},\sigma^{12}),
\end{equation}
with $\sigma^{\mu\nu}$ given by Eq.~(\ref{smunu}).

As expected, the helicity projector operators satisfy the following identities 
\cite{gre00,man86},
\begin{eqnarray}
[\Pi^\pm(\mathbf{p})]^2 &=& \Pi^\pm(\mathbf{p}), \\
\Pi^\pm(\mathbf{p})\Pi^\mp(\mathbf{p}) & = & 0, \\
\Pi^+(\mathbf{p}) + \Pi^-(\mathbf{p}) &=& \mathbb{1}.
\end{eqnarray}
Also, since the eigenvalues of $\sigma_{\mathbf{p}}$ are $\pm 1$, the eigenvalues of
$\Pi^\pm(\mathbf{p})$ are $0$ or $1$.

To carry over the remaining properties associated with $\Pi^\pm(\mathbf{p})$  
that are true for the standard Dirac equation to the present one 
(those properties based on Eqs.~(\ref{hu}) and (\ref{hv})),
we just need to prove that 
\begin{equation}
[\Lambda^+(\mathbf{p}),\Pi^\pm(\mathbf{p})] =  
[\Lambda^-(\mathbf{p}),\Pi^\pm(\mathbf{p})] = 0.
\label{lpcomuta}
\end{equation}
This is accomplished in appendix 
\ref{apC}. 

Equation (\ref{lpcomuta}) tells us that $\Lambda^\pm(\mathbf{p})$,  
$\Pi^\pm(\mathbf{p})$, $\sigma_{\mathbf{p}}$, 
and $\mathbf{p}=\mathbf{p}\mathbb{1}$ can all
be diagonalized together. Thus, we can choose the spinors $u_r(\mathbf{p})$ and
$v_r(\mathbf{p})$, which are eigenstates of $\Lambda^\pm(\mathbf{p})$, as follows, 
\begin{eqnarray}
\sigma_{\mathbf{p}}u_r(\mathbf{p})&=&(-1)^{r+1}u_r(\mathbf{p}), \label{hu} \\
\sigma_{\mathbf{p}}v_r(\mathbf{p})&=&(-1)^{r}v_r(\mathbf{p}). \label{hv}
\end{eqnarray}

The choices given by Eqs.~(\ref{hu}) and (\ref{hv}) are very useful when dealing with
the second quantization of the asymmetric Dirac equation \cite{rig23}, 
allowing a similar interpretation to the meaning of $u_r(\mathbf{p})$ and
$v_r(\mathbf{p})$ which is obtained when we second quantize the standard Dirac 
equation \cite{gre00,man86}. For instance, $u_1(\mathbf{p})$ represents a particle
with positive energy $E_{\mathbf{p}}^+$ and 
spin parallel to the direction of its motion given by $\mathbf{p}$, 
which we call a positive helicity state. For $u_2(\mathbf{p})$ we have a negative helicity state, with its spin antiparallel to its momentum $\mathbf{p}$. On the other hand,
$v_r(\mathbf{p})$ will be associated with antiparticles having positive energies 
given by $E_{\mathbf{p}}^-$, with the same momentum and helicity of the corresponding
particle represented by $u_r(\mathbf{p})$.

Finally, using Eqs.~(\ref{hpo}), (\ref{hu}) and (\ref{hv}) it can be shown that \cite{gre00,man86}
\begin{eqnarray}
\Pi^+(\mathbf{p})u_r(\mathbf{p}) &=& \delta_{1r}u_r(\mathbf{p}), \label{157}\\
\Pi^+(\mathbf{p})v_r(\mathbf{p}) &=& \delta_{2r}v_r(\mathbf{p}), \label{158}\\
\Pi^-(\mathbf{p})u_r(\mathbf{p}) &=& \delta_{2r}u_r(\mathbf{p}), \label{159}\\
\Pi^-(\mathbf{p})v_r(\mathbf{p}) &=& \delta_{1r}v_r(\mathbf{p}). \label{160}
\end{eqnarray}

We should also mention that since Eqs.~(\ref{ue}) and (\ref{ve}) are equal to
the corresponding ones related to the standard 
Dirac equation whenever $m=0$, we can express the helicity projection operators for a massless particle as \cite{gre00,man86}
\begin{equation}
\Pi^\pm(\mathbf{p}) = \frac{1}{2}(\mathbb{1} \pm \gamma^5).
\label{hpo0}
\end{equation}

\subsection{Spin projection operators}

Similarly to a standard Dirac particle, we have that only in the rest frame of 
the particle the spin component in an arbitrary direction is a good quantum number. 
However, it is possible to define in the rest frame and in a covariant way 
spin projection operators for an arbitrary quantization axis. From this definition,
we can go to any frame by implementing the appropriate Lorentz transformation.

For the asymmetric Dirac equation, the covariant spin projection operators in 
the rest frame along
the direction $\mathbf{\hat{n}}$ are 
\begin{equation}
\Pi^\pm(n) = \frac{1}{2}(\mathbb{1}\mp i\slashed n),
\label{spo}
\end{equation}
where
\begin{equation}
n^\mu = (0,\mathbf{\hat{n}}).
\label{nmu}
\end{equation}
Note that Eq.~(\ref{nmu}) implies that
\begin{eqnarray}
n^2=n^\mu n_\nu = -1 & \mbox{and} & np=n^\mu p_\mu = 0
\label{nn}
\end{eqnarray}
in all frames due to the invariance of scalar products. 
Indeed, in the rest frame Eq.~(\ref{nmu}) gives
$n^2=-|\mathbf{\hat{n}}|^2=-1$ and using that in the rest frame $p=(mc,0)$ we get
$np=0$.

Equation (\ref{spo}) is different from the 
standard Dirac particle spin projection operators,
namely, $\Pi^\pm_D(n)=(1/2)$ $(\mathbb{1}\pm \gamma^5\slashed n)$ \cite{man86}. 
The reason for this difference stems 
from Eq.~(\ref{dirA}) and the presence of the matrix $\gamma^5$ 
in the inertial term of that equation.

In the appendix 
\ref{apD} we prove that Eq.~(\ref{spo}) satisfies all the required 
properties of good spin projector operators \cite{man86}, in particular the fact that 
they commute with the energy projection operators for all $p$ satisfying Eq.~(\ref{nn}).

\section{Gordon's identities}

As we prove in the appendix 
\ref{apE}, if we use that
\begin{equation}
(p'+p)^\mu + i\sigma^{\mu\nu}(p'-p)_\nu = \slashed p' \gamma^\mu + \gamma^\mu \slashed p
\label{gor0}
\end{equation}
and Eqs.~(\ref{ue}), (\ref{ubare}), (\ref{ve}), and (\ref{vbare}),
we obtain the following four Gordon's identities,
\begin{eqnarray}
\overline{u}_s(\mathbf{p'})[(p'+p)^\mu &+& i\sigma^{\mu\nu}(p'-p)_\nu]u_r(\mathbf{p}) 
=  0, \label{gor1}\\ 
\overline{v}_s(\mathbf{p'})[(p'+p)^\mu &+& i\sigma^{\mu\nu}(p'-p)_\nu]v_r(\mathbf{p}) 
=  0, \label{gor2}\\ 
\overline{v}_s(\mathbf{p'})[(p'+p)^\mu &+& i\sigma^{\mu\nu}(p'-p)_\nu]u_r(\mathbf{p})  
\nonumber \label{gor3}\\
& = & 2imc\overline{v}_s(\mathbf{p'})\gamma^\mu\gamma^5u_r(\mathbf{p}), \\ 
\overline{u}_s(\mathbf{p'})[(p'+p)^\mu &+& i\sigma^{\mu\nu}(p'-p)_\nu]v_r(\mathbf{p})  
\nonumber \label{gor4}\\
& = & -2imc\overline{u}_s(\mathbf{p'})\gamma^\mu\gamma^5v_r(\mathbf{p}).
\end{eqnarray}

Furthermore, if repeat the steps leading to the previous Gordon's identities using 
Eq.~(\ref{gor0}) with $p \rightarrow -p$, we get these other four 
Gordon's identities,
\begin{eqnarray}
\overline{u}_s(\mathbf{p'})[(p'-p)^\mu &+& i\sigma^{\mu\nu}(p'+p)_\nu]u_r(\mathbf{p}) 
\nonumber \label{gor1a}\\
&=& -2imc\overline{u}_s(\mathbf{p'})\gamma^\mu\gamma^5u_r(\mathbf{p}) , \\ 
\overline{v}_s(\mathbf{p'})[(p'-p)^\mu &+& i\sigma^{\mu\nu}(p'+p)_\nu]v_r(\mathbf{p}) 
\nonumber \label{gor2a}\\
&=&  2imc\overline{v}_s(\mathbf{p'})\gamma^\mu\gamma^5v_r(\mathbf{p}), \\ 
\overline{v}_s(\mathbf{p'})[(p'-p)^\mu &+& i\sigma^{\mu\nu}(p'+p)_\nu]u_r(\mathbf{p})  
= 0, \label{gor3a}\\ 
\overline{u}_s(\mathbf{p'})[(p'-p)^\mu &+& i\sigma^{\mu\nu}(p'+p)_\nu]v_r(\mathbf{p})=0.
\label{gor4a}
\end{eqnarray}

\section{Plane wave solutions in the Dirac-Pauli representation}

Remembering that $\slashed p \slashed p = p^2=m^2c^2$, it is not difficult to see
that 
\begin{equation}
(\slashed p \mp imc \gamma^5)(\slashed p \mp imc \gamma^5) = 0.
\label{pp}
\end{equation}
Using Eq.~(\ref{pp}) and Eqs.~(\ref{ue}) and (\ref{ve}), we have the following free
particle solutions to the asymmetric Dirac equation in momentum space,
\begin{eqnarray}
u_r(\mathbf{p}) &=& N(\mathbf{p})(\slashed p -imc\gamma^5)v_{\overline{r}}(0), 
\label{udep}\\
v_r(\mathbf{p}) &=& N(\mathbf{p})(\slashed p +imc\gamma^5)u_{\overline{r}}(0), 
\label{vdep}
\end{eqnarray}
where
\begin{equation}
\overline{r} = 
\left\{
\begin{array}{ll}
1, \;\mbox{if}\;\; r=2, \\
2, \;\mbox{if}\;\; r=1,
\end{array}
\right.
\label{rbar2}
\end{equation}
and
\begin{equation}
N(\mathbf{p}) = \frac{1}{\sqrt{2mc(p^0+mc)}} =\frac{1}{\sqrt{2mE_{\mathbf{p}}+2m^2c^2}} 
\end{equation}
to guarantee that the normalization given by Eq.~(\ref{urvs}) is satisfied.

The zero momentum spinors $u_r(0)$ and $v_r(0)$ in the Dirac-Pauli representation (see appendix 
\ref{apD}) are, respectively, gi\-ven by Eqs. (\ref{u0}) and (\ref{v0}). And in Eqs.~(\ref{udep})
and (\ref{vdep}) we have $v_{\overline{r}}(0)$ and $u_{\overline{r}}(0)$ because, as 
a direct calculation in the present representation shows, we have
\begin{eqnarray}
-i\gamma^5v_{\overline{r}}(0) &=& \gamma^0v_{\overline{r}}(0)= u_r(0), \label{i30}\\
i\gamma^5u_{\overline{r}}(0) &=& \gamma^0u_{\overline{r}}(0) = v_r(0). \label{i40}
\end{eqnarray}
This ensures the correct ``initial condition'', namely, $u_r(\mathbf{p})$ and
$v_r(\mathbf{p})$ in Eqs.~(\ref{udep}) and (\ref{vdep}) become 
$u_r(0)$ and $v_r(0)$ when $\mathbf{p}=0$.

The corresponding plane wave solutions are, according to Eqs.~(\ref{ur}) and (\ref{vr}),
\begin{equation}
\Psi(x) = 
\left\{
\begin{array}{ll}
cte\; u_r(\mathbf{p})e^{i\kappa_\mu x^\mu} e^{-ip_\mu x^\mu/\hbar}, \\
cte\; v_r(\mathbf{p})e^{i\kappa_\mu x^\mu} e^{ip_\mu x^\mu/\hbar}, 
\end{array}
\right.
\label{urvr}
\end{equation}
and, if we explicitly compute Eqs.~(\ref{udep}) and (\ref{vdep}), we get 
\begin{eqnarray}
u_1(\mathbf{p}) &=& \frac{N(\mathbf{p})}{\sqrt{2}}
\left(\!\!
\begin{array}{c}
p^0+mc-ip^3 \\ p^2-ip^1 \\ p^3-i(p^0+mc) \\ p^1+ip^2
\end{array}
\!\!\right),\label{u1p}\\
u_2(\mathbf{p}) &=& \frac{N(\mathbf{p})}{\sqrt{2}}
\left(\!\!
\begin{array}{c}
-p^2-ip^1 \\ p^0+mc+ip^3\\ p^1-ip^2 \\ -p^3-i(p^0+mc) 
\end{array}
\!\!\right), \label{u2p} \\
v_1(\mathbf{p}) &=& \frac{N(\mathbf{p})}{\sqrt{2}}
\left(\!\!
\begin{array}{c}
p^2+ip^1 \\ p^0+mc-ip^3\\ p^1-ip^2 \\ -p^3+i(p^0+mc) 
\end{array}
\!\!\right),\label{v1p}\\
v_2(\mathbf{p}) &=& \frac{N(\mathbf{p})}{\sqrt{2}}
\left(\!\!
\begin{array}{c}
p^0+mc+ip^3 \\ -p^2+ip^1 \\ p^3+i(p^0+mc) \\ p^1+ip^2
\end{array}
\!\!\right).
\label{v2p}
\end{eqnarray}

It is worth noting that, similarly to the plane wave solutions of the standard Dirac 
equation in the Dirac-Pauli representation \cite{gre00,man86}, 
$u_r(\mathbf{p})$ and $v_r(\mathbf{p})$
as given above are not eigenstates of the helicity operator. A linear combination, though,
of $u_1(\mathbf{p})$ and $u_2(\mathbf{p})$ can always be found such that Eq.~(\ref{hu})
is satisfied and a linear combination of $v_1(\mathbf{p})$ and $v_2(\mathbf{p})$ can be
found such that Eq.~(\ref{hv}) is true.

There is a feature characteristic of Eqs.~(\ref{u1p})-(\ref{v2p}) that sets them apart
from the corresponding solutions of the standard Dirac equation \cite{gre00,man86}.
This is related to the fact that we cannot split them in an upper two-dimensional
spinor and a lower two-dimensional one,
where only the upper spinor or the lower one depends on $\mathbf{p}$. 
Also, looking at 
Eqs.~(\ref{u1p})-(\ref{v2p}) when $\mathbf{p}=0$ [cf. Eqs.~(\ref{u0})-(\ref{v0})], 
we realize that, in contradistinction to the plane wave solutions of the standard 
Dirac equation in the rest frame, both the upper and lower parts of $u_r(0)$ and $v_r(0)$
are not zero.

Before we finish this section, we will prove the following identities that 
will be helpful when we deal with the discrete symmetries associated with 
the asymmetric Dirac equation,
\begin{eqnarray}
u_r(\mathbf{p}) &=&  i\gamma^0\gamma^5u_r(\mathbf{-p}),  \label{i3}\\
v_r(\mathbf{p}) &=& - i\gamma^0\gamma^5v_r(\mathbf{-p}). \label{i4}
\end{eqnarray}

To prove Eq.~(\ref{i3}) we first realize that for $\mathbf{p}=0$ Eq.~(\ref{vdep}) can
be written as
\begin{equation}
v_{\overline{r}}(0) = \frac{1}{2}(\gamma^0 +i\gamma^5)u_{r}(0). 
\label{vde0}
\end{equation}
Inserting Eq.~(\ref{vde0}) into (\ref{udep}) we get after a little algebra
and using Eq.~(\ref{vdep}),
\begin{equation}
u_{r}(\mathbf{p}) = \frac{1}{2}\gamma^0 v_{\overline{r}}(-\mathbf{p})
- \frac{i}{2}\gamma^5 v_{\overline{r}}(\mathbf{p}).
\label{ubarp}
\end{equation}
If we now change $\mathbf{p}$ to $\mathbf{-p}$ in Eq.~(\ref{ubarp}) and left multiply
it by $i\gamma^5\gamma^0$, we obtain
\begin{equation}
i\gamma^5\gamma^0 u_{r}(\mathbf{-p}) = 
- \left[\frac{1}{2}\gamma^0 v_{\overline{r}}(-\mathbf{p})
- \frac{i}{2}\gamma^5 v_{\overline{r}}(\mathbf{p})\right]
= - u_{r}(\mathbf{p}).
\label{ubarp2}
\end{equation}
And since $\gamma^5\gamma^0=-\gamma^0\gamma^5$, Eq.~(\ref{ubarp2}) is equal to (\ref{i3}),
completing the proof.
The same reasoning can be used to prove Eq.~(\ref{i4}), exchanging the roles of 
Eqs.~(\ref{vdep}) and (\ref{udep}) while repeating the above logical steps.

\section{Minimal coupling prescription}
\label{mcp}

Using SI units and the metric signature of the present work, 
the electromagnetic minimal coupling prescription \cite{man86,gre95}
is implemented in the 
asymmetric Dirac equation by changing  all derivatives $\partial_\mu$
to
\begin{equation}
\partial_\mu \rightarrow D_\mu = \partial_\mu + \frac{iq}{\hbar}A_\mu.
\label{mc}
\end{equation}
The covariant four-vector potential is given by
\begin{equation}
A_\mu = \left( \frac{\varphi}{c}, -\mathbf{A} \right), 
\label{pres}
\end{equation}
where $\varphi$ and $\mathbf{A}=(A^1,A^2,A^3)$ are, respectively,
the electric and vector potentials describing an electromagnetic field.

Therefore, using Eq.~(\ref{mc}) in Eq.~(\ref{adeFinal2}) we obtain 
\begin{equation}
i\hbar \gamma^\mu D_\mu\Psi(x) - (imc\gamma^5 - 
\hbar \kappa_\mu \gamma^\mu)\Psi(x) = 0. 
\label{adeMC}
\end{equation}
The first term above, i.e., $i\hbar \gamma^\mu D_\mu\Psi(x)$, is formally
equal to the one coming from the standard Dirac equation when it is 
minimally coupled to the electromagnetic field. As such, many techniques used to solve
the Dirac equation in the presence of an external electromagnetic field can be 
carried over to solve the equivalent problem using the asymmetric Dirac equation.

A direct calculation shows that Eq.~(\ref{adeMC}) is covariant  
under proper Lorentz transformations if $\Psi$ transforms as given by Eq.~(\ref{MPsi}).
To arrive at that conclusion we should remember that 
after a proper Lorentz transformation $D_\mu$ transforms as 
a covariant four-vector since $\partial_\mu$ and $A_\mu$ are covariant 
four-vectors too.

Also, the gauge transformation
\begin{equation}
A_\mu(x) = \widetilde{A}_\mu(x) - \partial_\mu \chi(x)
\label{gaugeA}
\end{equation}
leads to
\begin{equation}
i\hbar \gamma^\mu \widetilde{D}_\mu\widetilde{\Psi}(x) - (imc\gamma^5 - 
\hbar \kappa_\mu \gamma^\mu)\widetilde{\Psi}(x) = 0.  
\label{gaugeB}
\end{equation}
In Eq.~(\ref{gaugeB}) we have 
$$
\widetilde{D}_\mu = \partial_\mu + \frac{iq}{\hbar}\widetilde{A}_\mu
$$
and
\begin{equation}
\Psi(x) = e^{i(q/\hbar)\chi(x)}\widetilde{\Psi}(x).
\label{gaugeC}
\end{equation}
The above argument proves that the asymmetric Dirac equation
minimally coupled  to the electromagnetic field is covariant after a local gauge
transformation, as given by Eq. (\ref{gaugeA}), 
if the wave function changes as prescribed by Eq. (\ref{gaugeC}).
Moreover, this covariance 
also implies that a local gauge transformation cannot be used to 
get rid of the constants $\kappa^\mu$.

\subsection{The Hydrogen atom}

Computing explicitly Eq.~(\ref{adeMC}) we get
\begin{equation}
i\hbar \gamma^\mu \partial_\mu\Psi(x) - mcB\Psi(x) - qA_\mu\gamma^\mu\Psi(x)= 0, 
\label{adeMC2}
\end{equation}
where $B$ is given by Eq.~(\ref{Bgeral2}). Using Eq.~(\ref{aDtoD}), carrying out the 
derivatives, and left multiplying
by $Ue^{-i\kappa_\mu x^\mu}$, we obtain
\begin{equation}
i\hbar \gamma^\mu \partial_\mu\Psi_D(x) - mc\Psi_D(x) - qA_\mu\gamma^\mu\Psi_D(x)= 0. 
\label{adeMC3}
\end{equation}
Equation (\ref{adeMC3}) is the standard Dirac equation after we apply to it the minimal coupling prescription. This means that we need to solve the standard Dirac equation to
obtain the solutions to the asymmetric Dirac one. 

Being more specific, any bound state problem involving regular matter 
associated with the asymmetric Dirac equation
will have the eigenvalues of the equivalent problem related to the standard Dirac
equation displaced by $-\hbar c \kappa^0$ [cf. the phase appearing in Eq.~(\ref{aDtoD})
and also Ref. \cite{rig22}]. The corresponding wave function (eigenvector) will be given 
by Eq.~(\ref{aDtoD}), where $\Psi_D(x)$ is the respective solution to 
the standard Dirac equation.

For the Hydrogen atom (static Coulomb problem), we have the eigenvalues
\begin{equation}
E_{n,j+1/2} = E_{n,j+1/2}^D - \hbar c \kappa^0, 
\end{equation}
with $E_{n,j+1/2}^D$ being the eigenvalues we obtain by solving the standard Dirac 
equation \cite{gre00},
%
\begin{equation}
E_{n,j+\frac{1}{2}}^{D}\hspace{-.1cm} =\hspace{-.08cm}  mc^2\hspace{-.1cm}
\left[ \hspace{-.08cm} 1 \hspace{-.08cm}+\hspace{-.1cm} \left(\hspace{-.1cm}
\frac{\alpha}{n-(j+\frac{1}{2})+\sqrt{(j+\frac{1}{2})^2-\alpha^2}}\hspace{-.1cm} 
\right)^{\hspace{-.1cm}2} \right]^{\hspace{-.1cm}-\hspace{-.05cm}\frac{1}{2}}
\hspace{-.1cm}.  
\label{D1}
\end{equation}
%
Here $n\geq 1$ is a positive integer and $j+1/2 = 1,2,\ldots,n$. The corresponding 
eigenvectors are according to Eq.~(\ref{aDtoD}),
\begin{equation}
\Psi_{n,j+1/2}(x) = e^{i\kappa_\mu x^\mu}\left(\frac{\mathbb{1}-i\gamma^5}{\sqrt{2}}\right)
\Psi_{D_{n,j+1/2}}(x),
\label{eigenvectors}
\end{equation}
with $\Psi_{D_{n,j+1/2}}(x)$ being the respective solution to the standard Dirac equation.
Note that in the particular case where $\kappa^0=mc/\hbar$ and $\bm{\kappa}=0$, the 
eigenvalues (\ref{D1}) are simply the ones coming from the usual Dirac equation with
the rest energy $mc^2$ subtracted from them.

Moreover, if we define the unitary operator
\begin{equation}
V(x) = e^{i\kappa_\mu x^\mu}\left(\frac{\mathbb{1}-i\gamma^5}{\sqrt{2}}\right),
\end{equation}
the expectation value of an observable $\hat{\mathcal{O}}$ 
according to Eq. (\ref{eigenvectors}), using the Dirac bra and ket 
notation, is
\begin{equation}
\langle \Psi_\lambda |\hat{\mathcal{O}} | \Psi_\lambda \rangle = 
\langle \Psi_{D_\lambda} |V^\dagger \hat{\mathcal{O}} V | \Psi_{D_\lambda} \rangle,
\label{obs0}
\end{equation}
where $\lambda$ denotes all relevant quantum numbers. 
Now, if 
\begin{equation}
V^\dagger \hat{\mathcal{O}} V = \hat{\mathcal{O}},
\end{equation}
i.e., if $V$ and $\hat{\mathcal{O}}$ commute, the predictions for the expectation value
of the observable $\hat{\mathcal{O}}$ are the same as the ones coming from the standard
Dirac equation. And since the non-trivial part of $V$ (the part not proportional to the 
identity matrix) is $\gamma^5$, whenever $[\hat{\mathcal{O}},\gamma^5]=0$ the asymmetric
and standard Dirac equations lead to the same predictions.

We can also build the asymmetric Dirac equation's observables in such a way
that we \textit{enforce} the two theories to give the same predictions. This can be done
by postulating that the any observable $\hat{\mathcal{O}}_D$ associated with the standard Dirac equation is mapped to an observable related to the asymmetric Dirac equation as follows,
\begin{equation}
\hat{\mathcal{O}} = V \hat{\mathcal{O}}_D V^\dagger.
\label{obs1}
\end{equation}
Inserting Eq.~(\ref{obs1}) into the right hand side of (\ref{obs0}) we get
\begin{equation}
\langle \Psi_\lambda |\hat{\mathcal{O}} | \Psi_\lambda \rangle = 
\langle \Psi_{D_\lambda} |\hat{\mathcal{O}}_D | \Psi_{D_\lambda} \rangle,
\end{equation}
which tells us that both theories yield the same predictions. The previous analysis is
valid, as we just showed, in the first quantization level. This is also  
true for the second quantized theory as we show in Ref. \cite{rig23}.

\section{Non-relativistic limit}

The non-relativistic limit of the asymmetric Dirac equation is most easily obtained by
using the transformation (\ref{aDtoD}) that connects it with the standard
Dirac equation and working in the Dirac-Pauli representation for the gamma matrices. 
As such, by using the non-relativistic limit of the latter equation we
can obtain the non-relativistic limit of the former via Eq.~(\ref{aDtoD}). 

According to Ref. \cite{gre00}, if we insert the ansatz 
\begin{equation}
\Psi_D(x) = e^{-i(mc/\hbar)x^0}\widetilde{\Psi}_D(x)
\label{DtildeD}
\end{equation}
into the minimally coupled standard Dirac equation we get
\begin{equation}
i\hbar \partial_t
\left(\hspace{-.1cm}
\begin{array}{c}
\widetilde{\varphi} \\
\widetilde{\chi}
\end{array}
\hspace{-.1cm}\right)
=
\left(\hspace{-.1cm}
\begin{array}{c}
c\bm{\tilde{\sigma}}\cdot \bm{\hat{\pi}}\widetilde{\chi}_{\!_D} \\
c\bm{\tilde{\sigma}}\cdot \bm{\hat{\pi}}\widetilde{\varphi}_{\!_D}
\end{array}
\hspace{-.1cm}\right)
+
qcA^0
\left(\hspace{-.1cm}
\begin{array}{c}
\widetilde{\varphi}_{\!_D} \\
\widetilde{\chi}_{\!_D}
\end{array}
\hspace{-.1cm}\right)
-2mc^2
\left(\hspace{-.1cm}
\begin{array}{c}
0 \\
\widetilde{\chi}_{\!_D}
\end{array}
\hspace{-.1cm}\right)\hspace{-.1cm},\label{nonRL}
\end{equation}
where
\begin{eqnarray}
\bm{\hat{\pi}} &=& \mathbf{\hat{p}} - q\mathbf{A}, \\
\mathbf{\hat{p}} &=& - i \hbar \nabla, \\
\bm{\tilde{\sigma}} &=& (\sigma_1,\sigma_2,\sigma_3), 
\label{pm} \\
\widetilde{\Psi}_D(x) & = & 
\left(
\begin{array}{c}
\widetilde{\varphi}_{\!_D}(x)\\
\widetilde{\chi}_{\!_D}(x)
\end{array}
\right). \label{twospinor}
\end{eqnarray}
In Eq.~(\ref{twospinor}) we should understand $\widetilde{\varphi}_{\!_D}$ and 
$\widetilde{\chi}_{\!_D}$ as two dimensional spinors and in Eq.~(\ref{pm})
$\sigma_j$ as the standard $2\times2$ Pauli matrices. 

In the non-relativistic limit
the particle's kinetic energy and its potential energy are small when
compared to its rest energy. This means that \cite{gre00}
\begin{eqnarray}
\left|i\hbar \partial_t \widetilde{\chi}_{\!_D}\right| &\ll& |mc^2\widetilde{\chi}_{\!_D}|, 
\label{kinetic}\\
\left|qcA^0 \widetilde{\chi}_{\!_D}\right| &\ll& |mc^2\widetilde{\chi}_{\!_D}|.
\label{potential}
\end{eqnarray}
Within this level of approximation, Eqs.~(\ref{kinetic}) and (\ref{potential}) imply
that the lower part of Eq.~(\ref{nonRL}) is solved if
\begin{equation}
\widetilde{\chi}_{\!_D} = \frac{\bm{\tilde{\sigma}}\cdot 
\bm{\hat{\pi}}}{2mc}\widetilde{\varphi}_{\!_D},
\label{phiLLchi}
\end{equation}
which, when inserted into the upper part of Eq.~(\ref{nonRL}) gives
\begin{eqnarray}
i\hbar \partial_t\widetilde{\varphi}_{\!_D}(x) &\hspace{-.2cm}=\hspace{-.2cm}& 
\left[\frac{(\bm{\tilde{\sigma}}\cdot \bm{\hat{\pi}})^2}{2m}
 + qcA^0\right]\widetilde{\varphi}_{\!_D}(x) \nonumber \\
&\hspace{-.2cm}=\hspace{-.2cm}& \left[ \frac{\bm{\hat{\pi}}^2}{2m} - \frac{q\hbar}{2m}\bm{\tilde{\sigma}} \cdot \mathbf{B}
+ qcA^0\right]\widetilde{\varphi}_{\!_D}(x). \label{paulieq}
\end{eqnarray}
Equation (\ref{paulieq}) is the Pauli equation written in SI units, the non-relativistic limit of the standard Dirac equation. Here 
$\mathbf{B} = \nabla \times \mathbf{A}$ and 
to obtain the last line of (\ref{paulieq}) we used
the identity $(\bm{\tilde{\sigma}}\cdot \bm{\hat{\pi}})^2 = \bm{\hat{\pi}}^2 
- q\hbar\bm{\tilde{\sigma}}\cdot \mathbf{B}$, a consequence of the algebra of the 
Pauli matrices \cite{gre00}.

Inserting Eq.~(\ref{DtildeD}) into (\ref{aDtoD}) we get
\begin{equation}
\Psi(x) = e^{i\kappa_\mu x^\mu -i\widetilde{m}x^0}\frac{1}{\sqrt{2}}
\left(
\begin{array}{c}
\widetilde{\varphi}_{\!_D} - i \widetilde{\chi}_{\!_D} \\
-i (\widetilde{\varphi}_{\!_D} + i \widetilde{\chi}_{\!_D})
\end{array}
\right).
\label{216}
\end{equation}
But in the non-relativistic limit $|\widetilde{\chi}_{\!_D}|\ll
|\widetilde{\varphi}_{\!_D}|$ because 
$\widetilde{\chi}_{\!_D}\approx$ $(v/2c) \widetilde{\varphi}_{\!_D}$ 
[cf. Eq.~(\ref{phiLLchi}) and Ref. \cite{gre00}]. Thus, in the 
non-relativistic approximation,
we can neglect the $\widetilde{\chi}_{\!_D}$ terms in Eq.~(\ref{216}),
\begin{equation}
\Psi(x) = e^{i\kappa_\mu x^\mu -i\widetilde{m}x^0}\frac{1}{\sqrt{2}}
\left(
\begin{array}{c}
\widetilde{\varphi}_{\!_D}(x)  \\
-i \widetilde{\varphi}_{\!_D}(x)
\end{array}
\right).
\label{217}
\end{equation}

If we now compute the time derivative of Eq.~(\ref{217}) and use (\ref{paulieq}) we get
\begin{eqnarray}
i\hbar \partial_t\Psi(x) &\hspace{-.2cm}=\hspace{-.2cm}& 
(\widetilde{m}-\kappa^0)\hbar c \Psi(x) 
+  e^{i\kappa_\mu x^\mu -i\widetilde{m}x^0} \nonumber \\
&\hspace{-.3cm}\times\hspace{-.5cm}& \left[ \frac{\bm{\hat{\pi}}^2}{2m} - \frac{q\hbar}{2m}\bm{\tilde{\sigma}} \!\cdot\! \mathbf{B}
+ qcA^0\right]\hspace{-.1cm}\frac{1}{\sqrt{2}}\hspace{-.1cm}
\left(\hspace{-.1cm}
\begin{array}{c}
\widetilde{\varphi}_{\!_D}(x)  \\
-i \widetilde{\varphi}_{\!_D}(x)
\end{array}
\hspace{-.1cm}\right) \nonumber \\
&=\hspace{-.1cm}& 
(\widetilde{m}-\kappa^0)\hbar c \Psi(x) 
+  e^{i\kappa_\mu x^\mu -i\widetilde{m}x^0} \nonumber \\
&\hspace{-.3cm}\times\hspace{-.5cm}& \left[ \frac{\bm{\hat{\pi}}^2}{2m} - \frac{q\hbar}{2m}\bm{\tilde{\sigma}}\! \cdot\! \mathbf{B}
+ qcA^0\right]
\hspace{-.1cm}e^{-i\kappa_\mu x^\mu +i\widetilde{m}x^0}\Psi(x), 
\nonumber \\
&& \label{paulieq2}
\end{eqnarray}
where the last line is obtained using Eq.~(\ref{217}).
With the aid of the identity,
\begin{equation}
\frac{\bm{\hat{\pi}}^2}{2m}[e^{i\bm{\kappa}\cdot \mathbf{r}}\Psi(x)]
= e^{i\bm{\kappa}\cdot \mathbf{r}}\frac{(\bm{\hat{\pi}}+\hbar\bm{\kappa})^2}{2m}
\Psi(x),
\end{equation}
we can finally rewrite Eq.~(\ref{paulieq2}) as
\begin{eqnarray}
i\hbar \partial_t\Psi(x) &\hspace{-.2cm}=\hspace{-.2cm}& 
(mc^2-c\hbar\kappa^0)\Psi(x) \nonumber \\
&&+\left[ \frac{(\bm{\hat{\pi}}+\hbar\bm{\kappa})^2}{2m} 
- \frac{q\hbar}{2m}\bm{\tilde{\sigma}} \cdot \mathbf{B}
+ qcA^0\right]\Psi(x). 
\nonumber \\
&& \label{paulieq3}
\end{eqnarray}
Equation (\ref{paulieq3}) is the non-relativistic approximation of the asymmetric
Dirac equation. It is the analog of the Pauli equation in the present context.

In the particular case where $\kappa^0=mc/\hbar$ and $\bm{\kappa}=0$,
Eq. (\ref{paulieq3}) becomes
\begin{equation}
i\hbar \partial_t\Psi(x) =
\left[ \frac{\bm{\hat{\pi}}^2}{2m} 
- \frac{q\hbar}{2m}\bm{\tilde{\sigma}} \cdot \mathbf{B}
+ qcA^0\right]\Psi(x), 
\label{paulieq4}
\end{equation}
which is formally equivalent to the Pauli equation (\ref{paulieq}).

On the other hand, in the general case in which $\bm{\kappa}\neq 0$, the following
transformation,
\begin{equation}
\Psi(x) = e^{i\kappa_\mu x^\mu-i\widetilde{m}x^0}\Psi_P(x),
\label{222}
\end{equation}
when inserted into Eq.~(\ref{paulieq3}) leads to
\begin{equation}
i\hbar\partial_t\Psi_P(x) =
\left[ \frac{\bm{\hat{\pi}}^2}{2m} 
- \frac{q\hbar}{2m}\bm{\tilde{\sigma}} \cdot \mathbf{B}
+ qcA^0\right]\Psi_P(x), 
\end{equation}
an equation formally equal to the Pauli equation. In this scenario,
it is more appropriate to consider $\Psi_P(x)$ as the non-relativistic 
approximation for the asymmetric Dirac equation wave function. Combining
Eqs.~(\ref{217}) and (\ref{222}) we immediately see that
\begin{equation}
\Psi_P(x) =\frac{1}{\sqrt{2}}
\left(
\begin{array}{c}
\widetilde{\varphi}_{\!_D}(x)  \\
-i \widetilde{\varphi}_{\!_D}(x)
\end{array}
\right).
\label{wfnrl}
\end{equation}

\section{Discrete symmetries}
\label{ds}

Our goal here is to build the parity, time reversal, and charge conjugation operators 
associated with the asymmetric Dirac equation. We will try to be as close as
possible to the way they are built for the standard Dirac equation. In addition to
that, we want the asymmetric Dirac equation to behave in the same way the usual Dirac equation behaves under the action of those discrete symmetry operations.

\subsection{Parity}

We want the asymmetric Dirac equation to be covariant under a parity operation.
The parity operation, or space inversion operation, changes the sign of the space
coordinates describing a given physical system,
\begin{equation}
x = (x^0, \mathbf{r}) \longrightarrow x' = (x^{0'}, \mathbf{r'}) = (x^0,-\mathbf{r}).
\label{si}
\end{equation}

The parity transformation (\ref{si}) can be studied along the same lines used to 
investigate proper Lorentz transformations. For the space inversion operation we
have
\begin{equation}
x'_\mu = \Lambda_{\!\mu}^{\;\nu} x_\nu,
\end{equation}
where
\begin{equation}
\Lambda_{\!0}^{\;0} = 1,\hspace{.1cm}  
\Lambda_{\!1}^{\;1} =\Lambda_{\!2}^{\;2} = \Lambda_{\!2}^{\;2} = -1, \hspace{.1cm}
\Lambda_{\!\mu}^{\;\nu} = 0,\hspace{.1cm} \mbox{if}\hspace{.1cm} \mu \neq \nu.
\label{pmunu}
\end{equation}
Note that $det(\Lambda)=-1$, which characterizes it as an improper Lorentz transformation.

The wave function changes according to
\begin{equation}
\Psi'(x') = M \Psi(x),
\end{equation}
where we assume that $M$ does not depend on the space-time coordinates,
\begin{equation}
\partial_\mu M = 0.
\label{delM}
\end{equation}

Repeating the steps given in Sec. \ref{rp}, and using Eq.~(\ref{delM}), the 
asymmetric Dirac equation is covariant under the parity operation if
\begin{eqnarray}
M^{-1}\gamma^\mu\Lambda_{\!\mu}^{\;\nu}M &=& e^{i\theta}\gamma^\nu,
\label{P1}\\
M^{-1}BM &=& e^{i\theta}B.
\label{P2}
\end{eqnarray}
Note that we are including an arbitrary global phase $\theta$ above. In this way,  Eqs.~(\ref{P1}) and (\ref{P2}) are the most general conditions guaranteeing the 
covariance of the asymmetric Dirac equation. If we set $\theta =0$, as we did when we 
studied its covariance under proper Lorentz transformations, we cannot obtain 
$M$ such that Eqs.~(\ref{P1}) and (\ref{P2}) are simultaneously satisfied.

Inspired by the parity operator of the Lorentz covariant Schr\"odinger equation \cite{rig22}
and by the fact that $B$ is a function of $\gamma^5$, $\gamma^\mu$, and $\kappa_\mu$, 
we define the parity operator for the asymmetric Dirac equation as
\begin{equation}
M = P_{\kappa} = \gamma^5 K_1 P.
\label{pk}
\end{equation}
Here $P$ is the parity operator of the standard Dirac equation \cite{gre00},
\begin{equation}
P = e^{i\varphi_{\!_P}}\gamma^0,
\end{equation}
with $\varphi_{\!_P} = 0, \pm \pi$, or $\pm \pi/2$. Note that 
these values for $\varphi_{\!_P}$ are obtained by postulating that four 
successive space inversions bring us back to the original spinor \cite{gre00}. 
We also have  
\begin{equation}
K_1 f(\kappa^\mu) K_1^\dagger = f(\kappa_\mu),
\label{k1}
\end{equation}
where $f(\kappa^\mu)$ is an arbitrary function of $\kappa^\mu$,
$K_1 = K_1^\dagger$, and $(K_1)^2=1$ \cite{rig22}. Remembering the
metric signature we are using in this work, the operator $K_1$ 
changes the sign of $\kappa^j$ while leaving
$\kappa^0$ unaltered. It is worth noting that $m$ is not changed 
by the action of $K_1$ since the dependence of $m$ on $\kappa^\mu$ is 
quadratic.  

As we show in the appendix 
\ref{apF}, the parity operator defined in Eq.~(\ref{pk})
satisfies Eqs.~(\ref{P1}) and (\ref{P2}) for $\theta=\pi$, which guarantees the 
covariance of the asymmetric Dirac equation under the parity operation.

\subsection{Time reversal}

Similarly to what we did for the parity operation, we want 
the asymmetric Dirac equation to be covariant after the time reversal operation.
The time reversal operation is an improper Lorentz transformation that changes the sign 
of the time coordinate associated with a given physical system,
\begin{equation}
x = (x^0, \mathbf{r}) \longrightarrow x' = (x^{0'}, \mathbf{r'}) = (-x^0,\mathbf{r}).
\label{tr}
\end{equation}

In other words, 
\begin{equation}
x'_\mu = \Lambda_{\!\mu}^{\;\nu} x_\nu,
\end{equation}
where
\begin{equation}
\Lambda_{\!0}^{\;0} = -1,\hspace{.1cm}  
\Lambda_{\!1}^{\;1} =\Lambda_{\!2}^{\;2} = \Lambda_{\!2}^{\;2} = 1, \hspace{.1cm}
\Lambda_{\!\mu}^{\;\nu} = 0,\hspace{.1cm} \mbox{if}\hspace{.1cm} \mu \neq \nu.
\label{trmunu}
\end{equation}

The wave function changes as 
\begin{equation}
\Psi'(x') = T_\kappa \Psi(x)
\end{equation}
after the time reversal operation, with $T_\kappa$ assumed to not dependent on $x^\mu$. 

Repeating the steps given in Sec. \ref{rp}, and now taking into account that
$T_\kappa$ should contain the complex conjugation operation ($T_\kappa$ does not 
commute with a complex number), 
the asymmetric Dirac equation is covariant under the time reversal operation if
\begin{eqnarray}
T_\kappa^{-1}i\gamma^\mu\Lambda_{\!\mu}^{\;\nu}T_\kappa &=& e^{i\theta}i\gamma^\nu,
\label{T1}\\
T_\kappa^{-1}BT_\kappa &=& e^{i\theta}B.
\label{T2}
\end{eqnarray}

As we prove in the appendix 
\ref{apF}, the operator
\begin{equation}
T_\kappa = \gamma^5 K_1 T,
\label{tk}
\end{equation}
where T is the time reversal operator of 
the standard Dirac equation \cite{gre00},
satisfies Eqs. (\ref{T1}) and (\ref{T2}) if $\theta=\pi$, guaranteeing the 
covariance of the asymmetric Dirac equation under the time reversal operation.

Note that in the Dirac-Pauli representation for the gamma matrices,
\begin{equation}
T = e^{i\varphi_{\!_T}}i\gamma^1\gamma^3K,
\label{Tdirac}
\end{equation}
with the complex conjugation operation $K=K^\dagger=K^{-1}$ defined by
\begin{equation}
KzK = z^*,
\end{equation}
where $z$ is an arbitrary complex number and 
$\varphi_{\!_T}$ a real number.

\subsection{Charge conjugation}

The asymmetric Dirac equation after we apply the minimal coupling prescription,
Eq. (\ref{adeMC2}), can be written as
\begin{equation}
i\hbar \slashed \partial\Psi(x) - i\hbar \widetilde{m}\gamma^5\Psi(x) 
+\hbar \slashed \kappa \Psi(x)- q\slashed A(x) \Psi(x)= 0. 
\label{adeMC4}
\end{equation}
The positive energy solutions to Eq.~(\ref{adeMC4}) are identified with particles of 
charge $q$ while the negative energy solutions are identified with antiparticles.

Since it is expected that antiparticles behave similarly to particles but 
with an opposite sign for their charge, we define the following asymmetric Dirac equation 
whose positive energy solutions have charge $-q$,
\begin{equation}
i\hbar \slashed \partial\Psi_c(x) - i\hbar \widetilde{m}\gamma^5\Psi_c(x) 
+\hbar \slashed \kappa \Psi_c(x)+ q\slashed A(x) \Psi_c(x)= 0. 
\label{adeMC5}
\end{equation}

As we will see, the charge conjugation operator is 
the operator that connects the solutions
of Eq.~(\ref{adeMC4}) to those of Eq.~(\ref{adeMC5}). In other words, it is 
the operator that transforms one equation into the other one \cite{gre00}. 
To obtain that operator, we first rewrite Eq.~(\ref{adeMC4}) by taking its complex conjugate,
where we have taken into account that $\widetilde{m}, \kappa^\mu$, and $A^\mu$ are all
real quantities,
\begin{equation}
[(i\hbar \partial_\mu -\hbar \kappa_\mu + q A_\mu)\gamma^{\mu^*} 
- i\hbar \widetilde{m}\gamma^{5^*}]\Psi^*(x) = 0. 
\label{adeMC6}
\end{equation}

If we now introduce an invertible matrix such that 
\begin{equation}
\Psi_c(x) = U_\kappa \Psi^*(x),
\label{psictopsi}
\end{equation}
we can rewrite Eq.~(\ref{adeMC6}) after left multiplying it by $U_\kappa$ as
\begin{eqnarray}
[i\hbar U_\kappa \gamma^{\mu^*} U_\kappa^{-1} \partial_\mu 
-\hbar U_\kappa \kappa_\mu \gamma^{\mu^*} U_\kappa^{-1} &&\nonumber \\
+ q A_\mu U_\kappa \gamma^{\mu^*} U_\kappa^{-1} 
- i\hbar \widetilde{m} U_\kappa \gamma^{5^*}U_\kappa^{-1}]\Psi_c(x) &=& 0. 
\label{adeMC7}
\end{eqnarray}
Note that we are already assuming that $U_\kappa$ commutes with complex numbers and
with $\widetilde{m}$ to write Eq.~(\ref{adeMC7}) as given above. The justification for
this rests in the fact that $U_\kappa$ does not depend on the complex conjugation 
operator $K$ and that $\widetilde{m}$ depends quadratically on $\kappa^\mu$, being  unaffected by $K_2$ as given by Eq.~(\ref{k2}). 

Comparing Eqs.~(\ref{adeMC5}) and (\ref{adeMC7}), they are the same up to an 
overall global phase if
\begin{eqnarray}
U_\kappa \gamma^{\mu^*} U_\kappa^{-1} & = & e^{i\theta}\gamma^\mu, 
\label{u1}\\
U_\kappa \gamma^{\mu^*}\kappa_\mu U_\kappa^{-1} & = & -e^{i\theta}\gamma^\mu\kappa_\mu, 
\label{u2}\\
U_\kappa \gamma^{5^*}U_\kappa^{-1} &=& e^{i\theta}\gamma^5.\label{u3}
\end{eqnarray}

As we prove in the appendix 
\ref{apF}, in the Dirac-Pauli representation for the 
gamma matrices the following operator satisfies Eqs.~(\ref{u1})-(\ref{u3}),
\begin{equation}
U_\kappa = C_\kappa \gamma^0,
\label{cg}
\end{equation}
where
\begin{eqnarray}
C_\kappa &=& K_2C, \label{ck}\\
C & = & e^{i\varphi_{\!_c}}i\gamma^2\gamma^0. \label{cdirac}
\end{eqnarray}
Here  $C_\kappa$ is an invertible operator, 
$C$ is the charge conjugation operator of the standard Dirac equation \cite{gre00},
with $\varphi_{\!_c}$ being a real number,  and 
\begin{equation}
K_2 f(\kappa^\mu) K_2^\dagger = f(-\kappa^\mu).
\label{k2}
\end{equation}
Note that the operator $K_2$ 
changes the sign of $\kappa^\mu$ and, as already anticipated,
$\widetilde{m}$ is not changed by its action since it 
depends quadratically on $\kappa^\mu$.  

Also, in the Dirac-Pauli representation for the gamma matrices we have 
$\gamma^0=\gamma^{0^T}$ and, thus, 
Eqs.~(\ref{psictopsi}) and (\ref{cg}) imply that
\begin{equation}
\Psi_c(x) = C_\kappa \overline{\Psi}^T\hspace{-.1cm}(x),
\label{psic}
\end{equation}
where $T$ means transposition. The operator $C_\kappa$ is the charge
conjugation operator for the asymmetric Dirac equation.

\subsection{$\kappa^\mu$ and the discrete symmetries}
\label{apFkmu}

In the asymmetric Dirac equation the four parameters $\kappa^0$ and 
$\bm{\kappa}=(\kappa^1,\kappa^2,\kappa^3)$ are ubiquitous. 
What we usually call the 
mass $m$ of a particle is a function of them [cf. Eqs.~(\ref{dispKG}) and (\ref{restmass})],
\begin{equation}
m = \frac{\hbar}{c}\sqrt{(\kappa^0)^2-|\bm{\kappa}|^2}.
\label{massF}
\end{equation}
We can understand the mass of a particle as originating from two different yet 
complementary aspects.
The first one is a ``time-like'' inertial contribution to the mass, 
given by $\kappa^0$, and the second one is a ``space-like'' inertial contribution,
given by $|\bm{\kappa}|$. The mass $m$ is proportional to $\sqrt{\kappa^\mu\kappa_\mu}$. 
Also, $\kappa^0$ is responsible for breaking the degeneracy of the energies associated 
with particles and antiparticles sharing the same wave number $\mathbf{k}$, while 
$\bm{\kappa}$ is responsible for breaking the degeneracy of their linear momenta
[cf. Eqs.~(\ref{ham2}) and (\ref{p2})].  

The parameters $\kappa^0,\kappa^1,\kappa^2$, and $\kappa^3$ 
are invariant under proper Lorentz transformations (boosts or
spatial rotations) \cite{rig22}. They are, roughly speaking, the analogs 
of the rest mass $m$ in the standard Dirac equation. 
However, $\kappa^\mu$ is not invariant under improper
Lorentz transformations, namely, parity and time reversal operations, and they are 
not invariant under the charge conjugation operation [cf. Eqs.~(\ref{k1}) and (\ref{k2})].
On the other hand, the mass $m$ is a relativistic invariant under proper and improper
Lorentz transformations. The latter can be seen by noting that $m$ depends quadratically
on $\kappa^\mu$ [cf. Eq.~(\ref{massF})].

Actually, looking carefully at the behavior of $\kappa^\mu$ under the discrete symmetry
operations studied in Sec. \ref{ds} and in the appendix 
\ref{apF},
we realize that the transformation rules of $\kappa^\mu$ under the 
parity, time reversal, and charge conjugation 
operations resemble the rules for the electromagnetic four-vector potential $A^\mu(x)$, 
namely \cite{gre00,gre95}, 
\begin{eqnarray}
A^\mu(x^0,\mathbf{r}) &\xrightarrow{\mbox{\tiny{parity}}}& A_\mu(x^0,-\mathbf{r}),\\
A^\mu(x^0,\mathbf{r}) &\xrightarrow{\mbox{\tiny{time reversal}}}& A_\mu(-x^0,\mathbf{r}),\\
A^\mu(x^0,\mathbf{r}) &\xrightarrow{\mbox{\tiny{charge conj.}}}& -A^\mu(x^0,\mathbf{r}).
\end{eqnarray}
Indeed, looking at Eqs.~(\ref{k1}) and (\ref{k2}) we get
\begin{eqnarray}
\kappa^\mu &\xrightarrow{\mbox{\tiny{parity}}}& \kappa_\mu, \label{parityF}\\
\kappa^\mu &\xrightarrow{\mbox{\tiny{time reversal}}}& \kappa_\mu, \label{trF}\\
\kappa^\mu &\xrightarrow{\mbox{\tiny{charge conj.}}}& -\kappa^\mu. \label{cgF}
\end{eqnarray}

This hybrid behavior of $\kappa^\mu$, being a Lorentz invariant when we have
proper Lorentz transformations and changing its sign according 
to Eqs.~(\ref{parityF}) and (\ref{trF}) when we deal with improper Lorentz transformations,
is a consequence of the assumptions we laid out when deriving the asymmetric Dirac equation. Being more specific,
$\kappa^\mu$ is invariant under proper Lorentz transformations because in this way we
guarantee that the asymmetric Dirac equation is covariant under those transformations and, at the same time, its free particle solutions have the same 
dispersion relations associated with 
the Lorentz covariant Schr\"odinger equation, where particles and antiparticles no longer
have the same energy for a given wave number \cite{rig22}. On the other hand,
$\kappa^\mu$ behaves as given by Eqs.~(\ref{parityF}) and (\ref{trF}) under improper 
Lorentz transformations because we imposed that
the asymmetric Dirac equation should be covariant after the parity or the time reversal operations. Similarly, by imposing that the asymmetric Dirac equation is covariant 
under the charge conjugation operation, we get Eq.~(\ref{cgF}). These latter assumptions
were brought to the present theory such that the asymmetric Dirac equation 
behaves in exactly the same way as the standard Dirac equation does
when subjected to those discrete symmetry operations. If we change the assumptions given by 
Eqs.~(\ref{parityF})-(\ref{cgF}), we can build a Lorentz covariant wave equation under
proper Lorentz transformation that responds differently under improper Lorentz 
transformations. We can build, for instance, 
a QED-like theory where we break the $CP$-symmetry from the start. See also
Refs. \cite{nis06,nis12,nis13,nis15} for other interesting approaches along this line.

In particular, we can build a theory violating the $P$-symmetry, the 
$C$-symmetry, or the $CP$-symmetry if we postulate a different transformation rule
for $\kappa^\mu$ after the parity and charge conjugation operations. 
To achieve that, instead of Eqs.~(\ref{parityF}) and (\ref{cgF}) we postulate that 
$\kappa^\mu$ transforms under those symmetry operations as follows,
\begin{equation}
\kappa^\mu \longrightarrow \kappa^\mu. \label{parityF2} 
\end{equation}
In other words, if $\kappa^\mu$ behaves as a strict scalar under the $C$ and $P$ 
operations, the asymmetric Dirac equation will not be covariant under 
those symmetry operations or combinations thereof.  This can be seen by 
noting that the assumptions given by Eqs.~(\ref{parityF}) and (\ref{cgF}) are crucial
to the proof that the asymmetric Dirac equation is covariant under the parity and charge conjugation operations (see appendix \ref{apF}).

The above analysis also 
expands and clarifies the discussion initiated in Ref. \cite{rig22} on
the possibility of ascribing to particles and antiparticles masses with opposite signs. 
What the discussion above tells us is that the mass $m$, as defined by Eq.~(\ref{massF}),
is positive for both particles and antiparticles, being invariant under proper and 
improper Lorentz transformations as well as under the charge conjugation operation.
Under the charge conjugation operation, though, we see that particles are associated with
$\kappa^\mu$ while antiparticles are associated with $-\kappa^\mu$. In the particular
and important case where $\kappa^j=0$ \cite{rig22}, we have $\kappa^0=mc/\hbar$ 
for particles and $\kappa^0=-mc\hbar$ for antiparticles, 
with both particles and antiparticles 
having a positive mass $m$ given by Eq.~(\ref{massF}). 
The implications of $\kappa^\mu$ having
different signs for particles and antiparticles may open the possibility 
to build a logical and coherent quantum theory of gravity in which particles
and antiparticles repel each other gravitationally, while particles attract particles and
antiparticles attract antiparticles, with both particles and antiparticles
having positive masses \cite{saf18,bon57,kow96,vil11,far18}. This comes about since
the gravitational interaction coupling constant is proportional to $\kappa^0$ and not to 
$m$ \cite{rig22}. However, further work is needed to complete this program 
(see also Sec. \ref{lf}).

Finally, we can understand the term 
\begin{equation}
-\overline{\Psi}(imc^2\gamma^5 - \hbar c\kappa_\mu\gamma^\mu)\Psi
\end{equation}
in the Lagrangian density (\ref{adeld2}), where $mc/\hbar = \sqrt{\kappa_\mu\kappa^\mu}$, 
as an alternative to the standard mass term
of the usual Dirac equation \cite{gre00,man86,gre95},
\begin{equation}
-mc^2\overline{\Psi}\Psi.
\end{equation}

These two terms, when added to the Lagrangian density
$i\hbar c \overline{\Psi}\gamma^\mu\partial_\mu$, 
are different ways of building a nonzero mass spin-1/2 theory from a massless one. 
The fundamental differences between the two approaches are: (1) the spinor 
$\Psi$ will have different transformation laws to guarantee the Lo\-ren\-tz covariance of 
the theory; (2) four parameters, $\kappa^0, \kappa^1,$ $\kappa^2, \kappa^3$,
instead of just one, $m$, to describe the inertial aspects of the particle; 
and (3) different energy-momentum expressions for particles and antiparticles.
And despite these differences, for interactions 
that respect the Lorentz symmetry we can build both theories
to yield the same predictions \cite{rig22,rig23}.\footnote{
The results
of Secs. \ref{cde}, \ref{mcp}, and \ref{ds} imply that the present
theory can be an alternative description of spin-1/2 particles,
for instance, electrons and positrons. By properly adjusting the free
parameters of the present theory, the standard Dirac equation and 
the asymmetric Dirac equation are equivalent descriptions of spin-1/2
particles.}

\section{Lagrangian formulation}
\label{lf}

We now consider $\Psi(x)$ and $\overline{\Psi}(x)$ as two independent fields and 
\begin{equation}
L = \int d^3x \mathcal{L}[\Psi(x),\overline{\Psi}(x),\partial_\mu \Psi(x),
\partial_\mu \overline{\Psi}(x)]
\end{equation}
the Lagrangian that completely characterizes the dynamics of those fields. 
The Lagrangian density $\mathcal{L}$ depends on the fields and on
their first derivatives. Here $d^3x = dx^1dx^2dx^3$ is the
infinitesimal spatial volume and the integrals above cover the entire space.
As usual, the fields and their derivatives are supposed to vanish at the boundaries of integration and the dimension of $\mathcal{L}$ is compatible with   
$L$ having the dimension of energy.

Noting that $d^4x = dx^0d^3x=cdtd^3x$ is the infinitesimal four-volume, the action is
defined as \cite{man86,gre95}
\begin{equation}
S = \int dt L = \frac{1}{c} \int d^4x 
\mathcal{L}(\Psi,\overline{\Psi},\partial_\mu \Psi,\partial_\mu \overline{\Psi}).
\label{acao}
\end{equation}
If the infinitesimal variation of the action vanishes, $\delta S = 0$, 
we obtain the following Euler-Lagrange equations \cite{man86,gre95},
\begin{eqnarray}
\frac{\partial \mathcal{L}}{\partial \Psi} = \partial_\mu\left( \frac{\partial \mathcal{L}}{\partial(\partial_\mu \Psi)}\right),\;\;\;
\frac{\partial \mathcal{L}}{\partial \overline{\Psi}} = \partial_\mu\left( \frac{\partial \mathcal{L}}{\partial(\partial_\mu \overline{\Psi})}\right).
\label{euler}
\end{eqnarray}

It is not difficult to see that the following Lagrangian density
gives the asymmetric Dirac equation and its adjoint 
when inserted into (\ref{euler}),
\begin{equation}
\mathcal{L} = i\hbar c  \overline{\Psi}(x) \gamma^\mu \partial_\mu \Psi(x)
-mc^2  \overline{\Psi}(x) B  \Psi(x),
\label{adeld}
\end{equation}
with $B$ given by Eq.~(\ref{Bgeral2}). Substituting $B$, and omitting the 
explicit dependence of the fields on $x$, we have
\begin{equation}
\boxed{\mathcal{L} = i\hbar c  \overline{\Psi} \gamma^\mu \partial_\mu \Psi
- \overline{\Psi} (imc^2\gamma^5-\hbar c \kappa_\mu\gamma^\mu)  \Psi,}
\label{adeld2}
\end{equation}
where we should not forget that $mc/\hbar=\sqrt{\kappa_\mu\kappa^\mu}$. 
A direct calculation shows that the Lagrangian density (\ref{adeld2}) is invariant under proper Lorentz transformations if $\Psi$ transforms according to Eq.~(\ref{Mfinite}),
with $M(x)$ satisfying Eqs.~(\ref{cond1}), (\ref{cond2a}), and (\ref{cond3}).\footnote{Note that (\ref{adeld2})
also leads to an action that is invariant under space-time translations.
This, together with the fact that  (\ref{adeld2})
is invariant under Lorentz transformations, implies that we have a theory
respecting the Poincar\'e symmetry \cite{gre95}.}

Since the time derivative appearing in Eq.~(\ref{adeld2}) are formally equal to 
the one in the Lagrangian density associated with the standard Dirac equation, we have
similar expressions for the conjugate momenta of the fields $\Psi$ and $\overline{\Psi}$,
\begin{eqnarray}
\Pi(x) = \frac{\partial\mathcal{L}}{\partial(\partial_t\Psi)}
=i\hbar\Psi^\dagger(x),\;\;
\overline{\Pi}(x) =\frac{\partial\mathcal{L}}{\partial(\partial_t\overline{\Psi})}
=0.
\end{eqnarray}

The canonical energy-momentum tensor for the asymmetric Dirac equation can be written
as \cite{man86,gre95}
\begin{equation}
\mathcal{T}^{\mu\nu} = \frac{\partial \mathcal{L}}{\partial(\partial_\mu\Psi)}\partial^\nu\Psi 
+\frac{\partial \mathcal{L}}{\partial(\partial_\mu\overline{\Psi})}\partial^\nu\overline{\Psi} 
- g^{\mu\nu}\mathcal{L},
\label{tmunu}
\end{equation}
leading to the following conserved Noether ``charges'',
\begin{equation}
cP^\mu = \int d^3x \mathcal{T}^{0\mu}(x).
\label{cPmu}
\end{equation}
In Eq.~(\ref{cPmu}) $cP^0$ is the Hamiltonian $H$ associated with the asymmetric Dirac 
Lagrangian and $P^j$ is its linear momentum.

A direct calculation gives
\begin{eqnarray}
\mathcal{T}^{\mu\nu} &=& i\hbar c \overline{\Psi}\gamma^\mu\partial^\nu\Psi 
-g^{\mu\nu}c\overline{\Psi}[i\hbar \slashed \partial -mcB]\Psi \nonumber \\
&=& i\hbar c \overline{\Psi}\gamma^\mu\partial^\nu\Psi, 
\label{tmunu2}
\end{eqnarray}
where we used that $\Psi$ is a solution to the asymmetric Dirac equation (\ref{adeFinal2})
to arrive at the last equality.

In the metric signature we have been using in this work, Eqs.~(\ref{cPmu}) and
(\ref{tmunu2}) imply that
\begin{eqnarray}
H =  i\hbar \int d^3x 
\Psi^\dagger(x)\partial_t\Psi(x), \label{ham}\\
\mathbf{P} = -i\hbar \int d^3x \Psi^\dagger(x)\nabla\Psi(x). \label{p}
\end{eqnarray}

When we second quantize this theory, we will have \cite{rig23}
\begin{eqnarray}
H &=&  \sum_{r=1}^2\int d^3p [(E_{\mathbf{p}}-\hbar c \kappa^0)
c^\dagger_r(\mathbf{p})c_r(\mathbf{p}) \nonumber \\
&&+(E_{\mathbf{p}}+\hbar c \kappa^0)d^\dagger_r(\mathbf{p})d_r(\mathbf{p})], \label{ham2}\\
\mathbf{P} &=& \sum_{r=1}^2\int d^3p [(\mathbf{p}-\hbar \bm{\kappa})
c^\dagger_r(\mathbf{p})c_r(\mathbf{p}) \nonumber \\
&&+(\mathbf{p}+\hbar \bm{\kappa})d^\dagger_r(\mathbf{p})d_r(\mathbf{p})], \label{p2}
\end{eqnarray}
with $c_r(\mathbf{p})$ and $d_r(\mathbf{p})$ being annihilation operators and 
$c^\dagger_r(\mathbf{p})$  and $d^\dagger_r(\mathbf{p})$ creation operators associated with
fermionic particles and antiparticles.

The conserved total angular momentum is computed in exactly the same way that is 
done for the standard Dirac Lagrangian \cite{man86,gre95} if we use $M^{ij}(x)$,
Eq.~(\ref{MmnGeral5}), instead of $S^{ij}=-(i/2)\sigma^{ij}$. The final result is
\begin{equation}
J^{k} = \frac{1}{c}\int d^3x[ x^i\mathcal{T}^{0j}(x)-x^j\mathcal{T}^{0i}(x)
+c\Pi(x)M^{ij}(x)\Psi(x)], \label{jk}
\end{equation}
with $i,j,k=1,2,3$ in cyclic order. Inserting Eqs.~(\ref{MmnGeral5}) and (\ref{tmunu2}) into (\ref{jk}) we get in vector notation,
\begin{eqnarray}
\mathbf{J} &=&  \int d^3x 
\Psi^\dagger(x)[\mathbf{r}\times (-i\hbar\nabla+\hbar\bm{\kappa})]\Psi(x) \nonumber \\
&&+ \int d^3x 
\Psi^\dagger(x)\left[\frac{\hbar}{2}\bm{\sigma}\right]\Psi(x).
\label{jvec}
\end{eqnarray}

Since $\Psi(x)$ is a solution to the asymmetric Dirac equation, we can insert the 
ansatz (\ref{aDtoD2}) into (\ref{jvec}). This gives
\begin{equation}
\mathbf{J} =  \int d^3x 
\widetilde{\Psi}^\dagger(x)\left[\mathbf{r}\times \mathbf{\hat{p}}  
+ \frac{\hbar}{2}\bm{\sigma}\right]\widetilde{\Psi}(x), 
\label{jvec2}
\end{equation}
where $\widetilde{\Psi}(x)$ does not depend of $\bm{\kappa}$ and 
$\mathbf{\hat{p}} = -i\hbar\nabla$.

Equation (\ref{jvec2}) tells us that the total angular momentum does not depend on
$\bm{\kappa}$ and that the first term in its right hand side is the orbital angular 
momentum while the second term is the intrinsic (spin) angular momentum associated with
the field $\Psi(x)$. Note that if we insert the ansatz (\ref{aDtoD2}) into  
Eq.~(\ref{p}), we will get a term depending on $\bm{\kappa}$, a feature that is reflected
in the second quantized expression for the linear momentum as given by Eq.~(\ref{p2}).
In other words, the asymmetry between particles and antiparticles manifests itself in 
different values for their energy and linear momentum  at a given wave number 
$\mathbf{k}=\mathbf{p}/\hbar$, while for the angular momentum the symmetry is still 
preserved (no $\bm{\kappa}$ dependence).

In addition to being invariant under space-time translations and spatial rotations,
the asymmetric Dirac Lagrangian density is also invariant under a global gauge 
transformation, $\Psi \rightarrow e^{i\epsilon}\Psi$, with $\epsilon$ an arbitrary real
number. The Noether theorem then leads to the following conserved charge \cite{man86,gre95},
\begin{equation}
Q = q \int d^3x \Psi^\dagger(x)\Psi(x), 
\label{Q}
\end{equation}
the following current,
\begin{equation}
\mathbf{j}(x) = cq\overline{\Psi}(x)\bm{\gamma} \Psi(x),
\end{equation}
with $\bm{\gamma}=(\gamma^1,\gamma^2,\gamma^2)$,
and the continuity equation
\begin{equation}
\partial_\mu s^\mu(x) = 0,
\end{equation}
where the four-current is
\begin{equation}
s^\mu(x) = (c\rho(x),\mathbf{j}(x)) = cq\overline{\Psi}(x)\gamma^\mu \Psi(x).
\end{equation}
The constant $q$ above is interpreted in the second quantization framework 
as the electric charge associated with the vacuum excitation created by 
$c^\dagger_r(\mathbf{p})$ and $-q$ the corresponding charge of the antiparticle 
created by $d^\dagger_r(\mathbf{p})$. Note that these results are consistent 
with the four-current obtained in Sec. \ref{cfc} by more elementary methods.

We can also apply the minimal coupling prescription to the Lagrangian density 
(\ref{adeld2}). This gives the following interaction Lagrangian density,
\begin{equation}
\mathcal{L}_{int} = -cq\overline{\Psi}(x)\gamma^\mu\Psi(x)A_\mu(x),
\label{adelint}
\end{equation}
which is formally equal to the one we obtain applying the minimal coupling prescription to
the standard Dirac equation.

On the other hand, we can phenomenologically model a non-electromagnetic interaction
using an external scalar potential, which is included in the Lagrangian density similarly to the way we add the standard mass term \cite{gre00}. In this 
case the interaction Lagrangian density becomes
\begin{equation}
\mathcal{L}_{ext} = -\mathcal{V}(x)\overline{\Psi}(x)\Psi(x),
\label{external}
\end{equation}
where $\mathcal{V}(x)$ is the potential energy associated to the interaction of 
the fermion field with the external field.  

In the theoretical framework of the asymmetric Dirac equation, in particular 
in the scenario where $\bm{\kappa}=0$ and $m=\hbar|\kappa^0|/c$
[see Eqs.~(\ref{dispKG}) and (\ref{restmass})], we can model the interaction of a fermion with a static gravitational field by setting
\begin{equation}
\mathcal{V}(x) = \frac{\hbar\kappa^0}{c}\varphi(x),
\label{extgravity}
\end{equation}
with $\varphi(x)$ being the gravitational potential related to the 
external field acting
on the fermion. Note that $\hbar\kappa^0/c$ is the mass of the 
particle if $\kappa^0>0$. If we insert Eq.~(\ref{extgravity}) into (\ref{external}) we
get
\begin{equation}
\mathcal{L}_{ext} = -\frac{\hbar\kappa^0}{c}\varphi(x)\overline{\Psi}(x)\Psi(x).
\label{external2}
\end{equation}

Now, looking at Eq.~(\ref{cgF}), we realize that the charge conjugation operation 
changes $\kappa^0$ to $-\kappa^0$ while $\overline{\Psi}(x)\Psi(x)$ is a scalar
under this symmetry operation \cite{rig23}. In this way, the interaction Lagrangian
density for an antiparticle interacting with this very same external gravitational field is 
\begin{equation}
\bar{\mathcal{L}}_{ext} = \frac{\hbar\kappa^0}{c}\varphi(x)\overline{\Psi}(x)\Psi(x).
\label{external3}
\end{equation}

Looking at Eqs.~(\ref{external2}) and (\ref{external3}), we notice that they differ 
by a minus sign. This means that if a particle is attracted by the 
gravitational field 
an antiparticle will be repelled by it. Particles and antiparticles can be modeled
to respond differently to a gravitational field within the framework of the 
asymmetric Dirac equation if we model the interaction with the gravitational field 
using Eqs.~(\ref{extgravity}) and (\ref{external2}). 

We can also make particles and antiparticles respond the same way to a gravitational
field if instead of Eq.~(\ref{external2}) we model their interaction as follows,
\begin{equation}
\tilde{\mathcal{L}}_{ext} = -\frac{\hbar|\kappa^0|}{c}\varphi(x)\overline{\Psi}(x)\Psi(x)
= -m\varphi(x)\overline{\Psi}(x)\Psi(x).
\label{external4}
\end{equation}
In other words, we now couple the fermion field with the gravitational field using
the magnitude of $\kappa^0$. The bottom line here is that in the theoretical framework
of the asymmetric Dirac equation, we can either assume that 
particles and antiparticles
attract each other gravitationally or that they repel each other without 
facing any logical contradiction. This comes about since in the two cases 
we do not need to assume $m$ to be negative. In the context of the 
asymmetric fields \cite{rig22,rig23}, the mass $m$ is always positive by 
construction [see Sec. \ref{secII}] and the way particles and
antiparticles couples to gravity is determined by $\kappa^0$ 
and not by $m$ \cite{rig22}.

We finish this section by noting that the Lagrangian density related to the wave
equation (\ref{dirA}) is 
\begin{equation}
\widetilde{\mathcal{L}}=i\hbar c \overline{\widetilde{\Psi}}\gamma^\mu \partial_\mu \widetilde{\Psi}
-imc^2\overline{\widetilde{\Psi}}\gamma^5\widetilde{\Psi}.
\end{equation}
The corresponding conserved Noether ``charges" related to its space-time 
invariance and the conserved electric char\-ge due to its global gauge invariance 
are exactly the ones given by 
Eqs.~(\ref{ham}), (\ref{p}), and (\ref{Q}), with $\Psi$ replaced by $\widetilde{\Psi}$, while its conserved total angular momentum is given by Eq.~(\ref{jvec2}).

\section{Conclusion}

We derived a first order spinorial wave equation whose free particle dispersion relations are equal to the dispersion relations associated with the Lorentz covariant Schr\"odinger equation \cite{rig22}. This latter equation is the relativistic analog 
of the standard non-relativistic Schr\"odinger equation, obtained by demanding 
Lorentz covariance and that
its wave function transforms under a proper Lorentz transformation 
according to the relativistic extension of the Schr\"odinger wave function's 
transformation law after a Galilean boost. 

In order to highlight the non-degenerate aspect of the energy-momentum relations 
for particles and antiparticles, we called it \textit{asymmetric Dirac equation}. 
We also determined how its solutions 
transform under proper Lorentz transformations by imposing that the  
asymmetric Dirac equation should have the same form after those transformations
(Lo\-rentz covariance). 

We then investigated the main similarities and 
differences between the present equation and the 
standard Dirac equation, providing 
a formal connection between the two equations. Throughout 
the development of the present theory, in particular when dealing with improper 
Lorentz transformations, we chose the path that led the asymmetric Dirac equation to 
behave as close as possible to the standard Dirac equation under the same circumstances
(physical conditions).
It turned out that we can build the present theory to either reproduce almost all predictions of the standard Dirac equation or we can follow a different yet logically 
consistent path, in which different predictions arise, such as a 
QED-like theory that violates the CP-symmetry from the start (see Sec. \ref{ds} and
appendix 
\ref{apF}).

We studied in details the plane wave solutions of the asymmetric Dirac equation
as well as its energy, helicity, and spin projection operators. We then obtained 
several Gordon's identities related to the asymmetric Dirac equation and we
introduced electromagnetic interactions via the minimal coupling prescription,
solving the respective Coulomb problem (hydrogen atom). 
We then determined the asymmetric Dirac equation non-relativistic limit, investigated 
its behavior under several discrete symmetry 
operations, and laid down the foundations of 
its classical field theory (Lagrangian formulation), 
preparing the ground to its second quantization that will be 
presented in Ref. \cite{rig23}. 

Finally, the present work shows that it is theoretically possible to construct 
a consistent Lorentz covariant spinorial wave equation using a more 
general transformation law for its wave function under a proper Lorentz transformation. 
Moreover, the free parameters of this wave equation can be adjusted to reproduce 
the predictions stemming from the standard Dirac equation. 
This latter fact is important since we must be able to 
predict all the experimentally validated results coming from the 
standard Dirac equation when applying the present theory in the domain
of validity of the standard Dirac equation. We showed that 
when working in this domain, the present theory can be adjusted to give exactly the same experimental predictions of the original Dirac equation.  
On the other hand, as a bonus, 
the present theory breaks the degeneracy between the energies of particles and 
antiparticles in its simplest version ($\kappa^0=mc/\hbar, \bm{\kappa}=0$) and 
also the degeneracy of their energies and momenta for a given wave number in its 
general version ($\kappa^0\neq |\bm{\kappa}|\neq 0$). The implications of the 
non-degenerate energy-momentum relations for particles and antiparticles at the second quantization level is more subtle and will be discussed elsewhere \cite{rig23}.

%

\begin{acknowledgments}
The author thanks the Brazilian agency CNPq
(National Council for Scientific and Technological Development) for partially
funding this research.
\end{acknowledgments}

\appendix


\section{Proof of Eq.~(\ref{PsiTildeTransf})}
\label{apA}

First, let us assume that 
\begin{equation}
\widetilde{\Psi}'(x')=\widetilde{S}\widetilde{\Psi}(x). 
\label{PsiTildeTransf2}
\end{equation}
Our goal is to prove that $\widetilde{S}=S$, with $S$ being the usual 
matrix that transforms a standard Dirac spinor after a proper Lorentz transformation.

Expressing the ansatz (\ref{aDtoD2}) in the primed reference frame we have
\begin{equation}
\Psi'(x') = e^{i\kappa_\mu x^{\mu'}}\widetilde{\Psi}'(x').
\label{aDtoD3}
\end{equation}
Also, using Eqs.~(\ref{Mfinite}) and (\ref{aDtoD2}) we get
\begin{equation}
\Psi'(x') = e^{i\kappa_\mu x^{\mu}}M(x)\widetilde{\Psi}(x).
\label{aDtoD4}
\end{equation}

Noting that the left hand sides of Eqs.~(\ref{aDtoD3}) and (\ref{aDtoD4}) are equal we
obtain
\begin{equation}
\widetilde{\Psi}'(x') = e^{i\kappa_\mu (x^{\mu}-x^{\mu'})}M(x)\widetilde{\Psi}(x).
\label{aDtoD5}
\end{equation}

Comparing Eqs.~(\ref{PsiTildeTransf2}) and (\ref{aDtoD5}) we finally arrive at
\begin{equation}
\widetilde{S} = e^{i\kappa_\mu (x^{\mu}-x^{\mu'})}M(x).
\label{Stilde}
\end{equation}

For an infinitesimal proper Lorentz transformation Eqs. (\ref{lt}) and (\ref{Minf}) 
imply, to first
order in $\epsilon_{\mu\nu}$, that Eq.~(\ref{Stilde}) becomes
\begin{equation}
\widetilde{S} = \mathbb{1} - \frac{i}{4}\epsilon_{\mu\nu}\sigma^{\mu\nu} 
+ \mathcal{O}(\epsilon_{\mu\nu}^2)= S.
\label{Stilde2}
\end{equation}
And since for any continuous symmetry operation we can build finite transformations 
from the infinitesimal ones, Eq. (\ref{Stilde2}) proves that $S=\widetilde{S}$ whether or not we have infinitesimal transformations. Finally, note that to arrive at 
Eq.~(\ref{Stilde2}), the term $i\epsilon_{\mu\nu} \kappa^\mu x^\nu$ coming from $M(x)$ was 
exactly canceled by the term $-i\epsilon_{\mu\nu}\kappa^\mu x^\nu$ coming from 
$e^{i\kappa_\mu (x^{\mu}-x^{\mu'})}$.

\section{The orthonormality relations and other identities}
\label{apB}

Let us start proving that the left hand side of Eq.~(\ref{o1}) is zero. If we left
multiply Eq.~(\ref{ve}) by $\overline{u}_r(\mathbf{-p})$ we get
\begin{equation}
\overline{u}_r(\mathbf{-p})(\gamma^0p_0 + \gamma^jp_j + imc\gamma^5)v_s(\mathbf{p})=0.
\label{1lo1}
\end{equation}
On the other hand, changing $\mathbf{p}$ to $\mathbf{-p}$ in Eq.~(\ref{ubare}) and right
multiplying it by $v_s(\mathbf{p})$ we obtain
\begin{equation}
\overline{u}_r(\mathbf{-p})(\gamma^0p_0 - \gamma^jp_j - imc\gamma^5)v_s(\mathbf{p})=0.
\label{2lo1}
\end{equation}
Summing Eqs.~(\ref{1lo1}) and (\ref{2lo1}) we get
\begin{equation}
p_0\overline{u}_r(\mathbf{-p})\gamma^0v_s(\mathbf{p})=0.
\end{equation}
Since by definition $p^0\neq 0$ we arrive at the desired expression,
\begin{equation}
u^\dagger_r(\mathbf{-p})v_s(\mathbf{p})=0 \Longleftrightarrow 
u^\dagger_r(\mathbf{p})v_s(\mathbf{-p})=0,
\label{o1p}
\end{equation}
after using Eq.~(\ref{uadj}).
To prove that the middle term of Eq. (\ref{o1}) is zero we just need to take the 
adjoint of (\ref{o1p}). 

Let us now prove that the left hand side of Eq.~(\ref{o2}) is zero. Left multiplying 
Eq.~(\ref{ue}) by $\overline{u}_r(\mathbf{p})\gamma^0$ we have
\begin{equation}
 \overline{u}_r(\mathbf{p})\gamma^0(\gamma^0p_0 + \gamma^jp_j 
- imc\gamma^5)u_s(\mathbf{p})=0.
\label{l1o2}
\end{equation}
On the other hand, right multiplying Eq.~(\ref{ubare}) by $\gamma^0u_s(\mathbf{p})$ and
using the anticommutation relations of $\gamma^\mu$ to bring $\gamma^0$ to the 
left we get
\begin{equation}
\overline{u}_r(\mathbf{p})\gamma^0(\gamma^0p_0 - \gamma^jp_j 
+ imc\gamma^5)u_s(\mathbf{p})=0.
\label{l2o2}
\end{equation}
Summing Eqs.~(\ref{l1o2}) and (\ref{l2o2}) and remembering that $(\gamma^0)^2$ $=\mathbb{1}$
and that $p_0\neq 0$ we obtain
\begin{equation}
\overline{u}_r(\mathbf{p})u_s(\mathbf{p})=0,
\label{o2p}
\end{equation}
which is what we wanted to prove.

Similarly we prove that the middle term of Eq.~(\ref{o2}) is zero. We just need to 
repeat the above proof using Eqs.~(\ref{ve}) and (\ref{vbare}) instead of 
(\ref{ue}) and (\ref{ubare}) and change appropriately the objects we left and 
right multiply them (use $v_r$ instead of $u_r$).

To prove that the left hand side of Eq.~(\ref{o3}) is zero, we left multiply 
Eq.~(\ref{ve}) by $\overline{u}_r(\mathbf{p})$ and subtract from it 
Eq.~(\ref{ubare}) right multiplied by $v_s(\mathbf{p})$. The middle term of 
Eq. (\ref{o3}) is zero since it is proportional to 
the adjoint of the left hand side term.

If we now left multiply Eq.~(\ref{ue}) by $\overline{u}_r(\mathbf{p})\gamma^0\gamma^5$ we
get
\begin{equation}
\overline{u}_r(\mathbf{p})\gamma^0\gamma^5 (\gamma^0p_0+\gamma^jp_j -imc\gamma^5)u_s(\mathbf{p})=0.
\label{l1o4}
\end{equation}
Right multiplying Eq.~(\ref{ubare}) by $\gamma^0\gamma^5u_s(\mathbf{p})$ and using the 
anticommutation rules involving $\gamma^5$ and $\gamma^\mu$ we obtain
\begin{equation}
\overline{u}_r(\mathbf{p})\gamma^0\gamma^5 (-\gamma^0p_0+\gamma^jp_j +imc\gamma^5)u_s(\mathbf{p})=0.
\label{l2o4}
\end{equation}
Subtracting Eq.~(\ref{l2o4}) from (\ref{l1o4}) and using that 
$\gamma^0\gamma^5=-\gamma^5\gamma^0$ and $(\gamma^0)^2=(\gamma^5)^2=\mathbb{1}$, we arrive
at 
\begin{equation}
\overline{u}_r(\mathbf{p})i\gamma^5u_s(\mathbf{p}) = 
\frac{mc}{p_0}u_r^\dagger(\mathbf{p})u_s(\mathbf{p}).
\label{l3o4}
\end{equation}
after using Eq.~(\ref{uadj}). And with the help of Eq.~(\ref{urvs}) we  
prove that the left hand side of Eq.~(\ref{o4}) is $\delta_{rs}$. To prove that 
the middle term of (\ref{o4}) is also $\delta_{rs}$, we proceed similarly, using 
Eq.~(\ref{ve}) left multiplied by $\overline{v}_r(\mathbf{p})\gamma^0\gamma^5$ and Eq.~(\ref{vbare}) right multiplied by $\gamma^0\gamma^5v_s(\mathbf{p})$.

The first resolution of the identity, Eq.~(\ref{complete1}), can be proven by acting 
on the four base vectors $u_s(\mathbf{p})$ and $v_s(\mathbf{p})$, with $s=1,2$, and 
verifying that we obtain the expected results, namely, 
$\mathbb{1}u_s(\mathbf{p})=u_s(\mathbf{p})$ and $\mathbb{1}v_s(\mathbf{p})=v_s(\mathbf{p})$.

For instance, if we use Eqs.~(\ref{o3}) and (\ref{o4}) we get
\begin{eqnarray}
\mathbb{1}u_s(\mathbf{p}) &=& \sum_{r=1}^2
[u_r(\mathbf{p})\overline{u}_r(\mathbf{p})i\gamma^5u_s(\mathbf{p}) \nonumber \\
&& - v_r(\mathbf{p})\overline{v}_r(\mathbf{p})i\gamma^5u_s(\mathbf{p})] \nonumber \\
&=& \sum_{r=1}^2
[u_r(\mathbf{p})\delta_{rs}-0] = u_s(\mathbf{p}). 
\end{eqnarray}
In an analogous way we show that $\mathbb{1}v_s(\mathbf{p})=v_s(\mathbf{p})$.

To prove the second resolution of the identity, Eq.~(\ref{complete2}), we take the
adjoint of (\ref{complete1}) and then left and right multiply it by $\gamma^0$. Since
$\gamma^0\mathbb{1}^\dagger\gamma^0=\mathbb{1}$, we get
\begin{eqnarray}
\mathbb{1} &=& \sum_{r=1}^2[\gamma^0(-i\gamma^5)\gamma^0u_r(\mathbf{p})
u_r^\dagger(\mathbf{p})\gamma^0\nonumber \\
&& - \gamma^0(-i\gamma^5)\gamma^0v_r(\mathbf{p})v_r^\dagger(\mathbf{p})\gamma^0] 
\nonumber \\
&=&\sum_{r=1}^2[i\gamma^5u_r(\mathbf{p})\overline{u}_r(\mathbf{p})\nonumber - i\gamma^5v_r(\mathbf{p})\overline{v}_r(\mathbf{p})], 
\end{eqnarray}
which is exactly Eq.~(\ref{complete2}) we wanted to prove.

Looking at Eqs.~(\ref{i1}) and (\ref{i2}), we realize that the latter is obtained from
the former by taking its adjoint. Therefore, we just need to prove Eq.~(\ref{i1}). 
If we change $\mathbf{p}$ to $\mathbf{-p}$ in Eq.~(\ref{ubare}), right multiply it
by $\gamma^5v_r(\mathbf{p})$, and use that $\gamma^5$ anticommutes with all $\gamma^\mu$,
we obtain
\begin{equation}
\overline{u}_s(\mathbf{-p})\gamma^5(-\gamma^0p_0 + \gamma^jp_j - imc\gamma^5)v_r(\mathbf{p})=0.
\label{1i1}
\end{equation}
Similarly, multiplying Eq.~(\ref{ve}) by $\overline{u}_s(\mathbf{-p})\gamma^5$ at the left
we get
\begin{equation}
\overline{u}_s(\mathbf{-p})\gamma^5(\gamma^0p_0 + \gamma^jp_j + imc\gamma^5)v_r(\mathbf{p})=0.
\label{2i1}
\end{equation}
Subtracting Eq.~(\ref{1i1}) from (\ref{2i1}), using Eq.~(\ref{uadj}), employing that 
$(\gamma^5)^2=\mathbb{1}$, and that $p_0=p^0$, 
we arrive at the desired expression,
\begin{equation}
\overline{u}_s(\mathbf{-p})v_r(\mathbf{p}) = -\frac{ip^0}{mc}
u^\dagger_s(\mathbf{-p})\gamma^5v_r(\mathbf{p}).
\label{i1p}
\end{equation}

\section{Proof of Eq.~(\ref{lpcomuta})}
\label{apC}

Using Eqs.~(\ref{epo}) and (\ref{hpo}) we have
\begin{eqnarray}
[\Lambda^+(\mathbf{p}), \Pi^\pm(\mathbf{p})] &=& 
\frac{\pm1}{4imc}[\gamma^5\slashed p, \sigma_{\mathbf{p}}] \nonumber \\
&=& \frac{\pm1}{4imc} \{ \gamma^5[\slashed p,  \sigma_{\mathbf{p}}]
+ [\gamma^5, \sigma_{\mathbf{p}}]\slashed p \} \nonumber \\
&=& \frac{\pm1}{4imc} [\gamma^5, \sigma_{\mathbf{p}}]\slashed p,
\label{apc1}
\end{eqnarray}
where we used that $[\slashed p, \sigma_{\mathbf{p}}] = 0$ \cite{gre00}.

If we now use Eq.~(\ref{sigmap}) and that 
\begin{equation}
\sigma^{ij}=-\gamma^0\gamma^5\gamma^k,
\label{sij}
\end{equation}
with
$i,j,k=1,2,3$ in cyclic order \cite{man86}, we obtain
\begin{eqnarray}
[\gamma^5, \sigma_{\mathbf{p}}] & = & 
\frac{1}{|\mathbf{p}|}[\gamma^5,\gamma^0\gamma^5\gamma^kp_k] \nonumber \\
&=&\frac{1}{|\mathbf{p}|}(\gamma^5\gamma^0\gamma^5\gamma^kp_k 
-\gamma^0\gamma^5\gamma^k\gamma^5p_k ) \nonumber \\
&=& \frac{1}{|\mathbf{p}|} (-\gamma^0\gamma^kp_k +\gamma^0\gamma^kp_k) = 0,
\label{apc2}
\end{eqnarray}
where we used that $\gamma^\mu\gamma^5=-\gamma^5\gamma^\mu$ and 
$\gamma^5\gamma^5=\mathbb{1}$ to obtain the last line.

Using Eqs.~(\ref{apc1}) and (\ref{apc2}) we get 
$[\Lambda^+(\mathbf{p}), \Pi^\pm(\mathbf{p})]=0$. In an analogous way we prove that
$[\Lambda^-(\mathbf{p}), \Pi^\pm(\mathbf{p})]=0$.

\section{The spin projector operators}
\label{apD}

Let us start proving that the spin projector operators commute with the energy 
projection operators. Using Eqs.~(\ref{epo}) and (\ref{spo}) we get
\begin{equation}
[\Lambda^+(\mathbf{p}), \Pi^\pm(n)] = \frac{\mp 1}{4mc}[\gamma^5\slashed p, \slashed n].
\end{equation}
But
\begin{equation}
[\gamma^5\slashed p, \slashed n]  = \gamma^5\slashed p \slashed n  
- \slashed n\gamma^5\slashed p = \gamma^5 (\slashed p \slashed n + \slashed n\slashed p).
\end{equation}
However, since $pn=0$ [cf. Eq.(\ref{nn})] we have $\slashed p \slashed n = -\slashed n\slashed p$ \cite{man86} and thus $[\gamma^5\slashed p, \slashed n]  =0$. This proves
that
\begin{equation}
[\Lambda^+(\mathbf{p}), \Pi^\pm(n)] = 0
\end{equation}
and similarly we can prove that $[\Lambda^-(\mathbf{p}), \Pi^\pm(n)] = 0$. It is worth
mentioning that the standard Dirac spin projector operators $\Pi^\pm_D(n)$
do not commute with $\Lambda^\pm(\mathbf{p})$ as given by Eq.~(\ref{epo}).

If we now use that $\slashed n \slashed n = n^2 =-1$, we obtain 
\begin{eqnarray}
[\Pi^\pm(n)]^2 &=& \Pi^\pm(n), \\
\Pi^\pm(n)\Pi^\mp(n) & = & 0, \\
\Pi^+(n) + \Pi^-(n) &=& \mathbb{1},
\end{eqnarray}
the expected properties of a complete set of projector operators.

What remains to be done is to check that $\Pi^\pm(n)$ has the expected properties
in the particle's rest frame. To simplify the following calculations, 
we need to choose a particular representation for the gamma matrices. 
From now on we will be working 
with the Dirac-Pauli representation of the gamma matrices, which is defined as 
follows \cite{gre00,man86},
\begin{eqnarray}
\gamma^0 =
\left(\!\!
\begin{array}{rr}
\mathbf{1} & \mathbf{0} \\
\mathbf{0} & -\mathbf{1}
\end{array}
\!\!\right),\;\;
\gamma^j =
\left(\!\!
\begin{array}{cl}
\mathbf{0} & \sigma_j \\
-\sigma_j & \mathbf{0}
\end{array}
\!\!\right), 
\label{eqd7}
\end{eqnarray}
where $\mathbf{1}$ and $\mathbf{0}$ are $2\times2$ matrices,
\begin{eqnarray}
\mathbf{1} =
\left(\!\!
\begin{array}{lr}
1 & 0 \\
0 & 1
\end{array}
\!\!\right),\;\;
\mathbf{0} =
\left(\!\!
\begin{array}{lr}
0 & 0 \\
0 & 0
\end{array}
\!\!\right), 
\label{eqd8}
\end{eqnarray}
and the $2\times2$ Pauli matrices are
\begin{eqnarray}
\sigma_1 =
\left(\!\!
\begin{array}{lr}
0 & 1 \\
1 & 0
\end{array}
\!\!\right),\;\;
\sigma_2 =
\left(\!\!
\begin{array}{lr}
0 & -i \\
i & 0
\end{array}
\!\!\right),\;\; 
\sigma_3 =
\left(\!\!
\begin{array}{lr}
1 & 0 \\
0 & -1
\end{array}
\!\!\right). 
\label{eqd9}
\end{eqnarray}

Using Eqs.~(\ref{eqd7})-(\ref{eqd9}) and Eqs.~(\ref{gamma5})-(\ref{smunu}) we obtain
\begin{equation}
\gamma^5 =
\left(\!\!
\begin{array}{lr}
\mathbf{0} & \mathbf{1} \\
\mathbf{1} & \mathbf{0}
\end{array}
\!\!\right), \;\; 
\sigma^{ij} =
\left(\!\!
\begin{array}{ll}
\sigma_k & 0 \\
0 & \sigma_k
\end{array}
\!\!\right), 
\label{eqd10}
\end{equation}
with $i,j,k=1,2,3$ in cyclic order.

In the particle's rest frame Eq.~(\ref{nmu}) implies that Eq.~(\ref{spo}) can be written as
\begin{equation}
\Pi^\pm(\mathbf{\hat{n}}) = \frac{1}{2}[\mathbb{1}\pm \bm{\sigma}\cdot\mathbf{\hat{n}}
(i\gamma^0\gamma^5)].
\label{eqd11}
\end{equation}
To arrive at Eq.~(\ref{eqd11}) we used that $(\gamma^5)^2=(\gamma^0)^2=\mathbb{1}$,
the anticommutation properties involving $\gamma^\mu$ and $\gamma^5$, and
Eq.~(\ref{sij}).
In the rest frame, Eqs.~(\ref{ue}) and (\ref{ve}) give
\begin{eqnarray}
u_r(0) = i\gamma^0\gamma^5 u_r(0), & v_r(0) = -i\gamma^0\gamma^5 v_r(0).  
\label{eqd12}
\end{eqnarray}

Using Eqs.~(\ref{eqd11}) and (\ref{eqd12}) we obtain
\begin{eqnarray}
\Pi^\pm(\mathbf{\hat{n}}) u_r(0) &=& \frac{1}{2}(\mathbb{1}\pm \bm{\sigma}\cdot\mathbf{\hat{n}})u_r(0), \label{eqd13} \\
\Pi^\pm(\mathbf{\hat{n}}) v_r(0) &=& \frac{1}{2}(\mathbb{1}\mp \bm{\sigma}\cdot\mathbf{\hat{n}})v_r(0). \label{eqd14}
\end{eqnarray}
The operators $\frac{1}{2}(\mathbb{1}\pm \bm{\sigma}\cdot\mathbf{\hat{n}})$ are exactly
the spin projection operators along the direction $\mathbf{\hat{n}}$ of 
the non-relativistic quantum mechanics ($v/c\ll 1$). 

In the Dirac-Pauli representation
of the gamma matrices, Eqs.~(\ref{ue}) and (\ref{ve})
imply that when $\mathbf{p}=0$ (particle's rest frame) we have, 
\begin{eqnarray}
u_1(0) = \frac{1}{\sqrt{2}}
\left(\!\!
\begin{array}{c}
1 \\ 0 \\ -i \\ 0
\end{array}
\!\!\right),\,\,
u_2(0) = \frac{1}{\sqrt{2}}
\left(\!\!
\begin{array}{c}
0 \\ 1 \\ 0 \\ -i
\end{array}
\!\!\right), \label{u0} \\
v_1(0) = \frac{1}{\sqrt{2}}
\left(\!
\begin{array}{c}
0 \\ 1 \\ 0 \\ i
\end{array}
\!\right),\,\,
v_2(0) = \frac{1}{\sqrt{2}}
\left(\!
\begin{array}{c}
1 \\ 0 \\ i \\ 0
\end{array}
\!\right).
\label{v0}
\end{eqnarray}
The normalization chosen above guarantees the validity of Eq.~(\ref{urvs}).

We now orient, for simplicity of calculation and without loss of generality,
$\mathbf{\hat{n}}$ along the $z$-direction ($x^3$-direction), i.e., we set
$\mathbf{\hat{n}}=\mathbf{\hat{z}}=(0,0,1)$. In this scenario Eqs.~(\ref{u0})
and (\ref{v0}) give
\begin{eqnarray}
(\bm{\sigma}\cdot\mathbf{\hat{n}})u_r(0)&=&(-1)^{r+1}u_r(0), \label{su0} \\
(\bm{\sigma}\cdot\mathbf{\hat{n}})v_r(0)&=&(-1)^{r}v_r(0), \label{sv0}
\end{eqnarray}
the analogs of Eqs.~(\ref{hu}) and (\ref{hv}).

Finally, using Eqs.~(\ref{su0}) and (\ref{sv0}) in 
Eqs.~(\ref{eqd13}) and (\ref{eqd14}) we get
\begin{eqnarray}
\Pi^+(\mathbf{\hat{n}}) u_r(0) &=& \delta_{1r}u_r(0), \label{eqd19}\\ 
\Pi^+(\mathbf{\hat{n}}) v_r(0) &=& \delta_{1r}v_r(0), \label{eqd20}\\
\Pi^-(\mathbf{\hat{n}}) u_r(0) &=& \delta_{2r}u_r(0), \label{eqd21}\\
\Pi^-(\mathbf{\hat{n}}) v_r(0) &=& \delta_{2r}v_r(0). \label{eqd22}
\end{eqnarray}
The above relations tell us that $u_1(0)$ and $v_1(0)$ have spins pointing along
the direction of $\mathbf{\hat{n}}$ while $u_2(0)$ and $v_2(0)$ have spins 
oriented in the opposite direction.

Note that due to the different ordering in sign
that we see at the right hand side of Eqs.~(\ref{eqd13}) and (\ref{eqd14}), namely,
$\pm$ and $\mp$, Eqs.~(\ref{eqd20}) and (\ref{eqd22}) are formally different from Eqs.~(\ref{158}) and (\ref{160}).  

\section{Proof of the Gordon's identities}
\label{apE}

Let us start proving Eq.~(\ref{gor0}). If we use Eq.~(\ref{smunu}), the left hand side (lhs)
of Eq.~(\ref{gor0}) can be written after a little algebra as
\begin{equation}
lhs = p^{\mu'} + p^\mu 
+ \frac{1}{2}(-\gamma^\mu\gamma^\nu p'_\nu + \slashed p' \gamma^\mu) 
+ \frac{1}{2}(-\gamma^\nu\gamma^\mu p_\nu + \gamma^\mu \slashed p). 
\end{equation}
If we now use Eq.~(\ref{c1}) to anticommute $\gamma^\mu\gamma^\nu$ and 
$\gamma^\nu\gamma^\mu$ above we get
\begin{equation}
lhs = \slashed p' \gamma^\mu  + \gamma^\mu \slashed p, 
\end{equation}
which is the right hand side (rhs) of Eq.~(\ref{gor0}).

To prove Eq.~(\ref{gor1}) we insert Eq.~(\ref{gor0}) into the lhs of (\ref{gor1}) and
then use Eqs.~(\ref{ue}) and (\ref{ubare}). Proceeding as described, we arrive at
\begin{equation}
lhs = imc\overline{u}_s(\mathbf{p'})[\gamma^5\gamma^\mu +\gamma^\mu\gamma^5]
u_r(\mathbf{p}) = 0,
\end{equation}
since $\gamma^5$ and $\gamma^\mu$ anticommute. This proves Eq.~(\ref{gor1}) and 
analogously we prove Eq.~(\ref{gor2}) repeating the steps described above and 
using Eqs.~(\ref{ve}) and (\ref{vbare}) instead of Eqs.~(\ref{ue}) and (\ref{ubare}).

Equation (\ref{gor3}) can be proved after we insert Eq.~(\ref{gor0}) into its lhs and
then use Eqs.~(\ref{ue}) and (\ref{vbare}) to get
\begin{eqnarray}
lhs &=& imc\overline{v}_s(\mathbf{p'})[-\gamma^5\gamma^\mu +\gamma^\mu\gamma^5]
u_r(\mathbf{p}) \nonumber \\
&=&2imc\overline{v}_s(\mathbf{p'})\gamma^\mu\gamma^5u_r(\mathbf{p}),
\end{eqnarray}
which is equal to the rhs of Eq.~(\ref{gor3}). We obtain Eq.~(\ref{gor4}) in the same
way, using, at the end of the proof, Eqs.~(\ref{ubare}) and (\ref{ve}) instead
of Eqs.~(\ref{ue}) and (\ref{vbare}).

Finally, the proofs leading to Eqs.~(\ref{gor1a})-(\ref{gor4a}) can be readily obtained by
slightly changing the four proofs given above if we  insert
Eq.~(\ref{gor0}) with $p\rightarrow -p$ at the lhs of Eqs.~(\ref{gor1a})-(\ref{gor4a})
and then use the appropriate pair of equations 
among Eqs.~(\ref{ue}), (\ref{ubare}), (\ref{ve}), and (\ref{vbare}).

\section{More on discrete symmetries}
\label{apF}

\subsection{Parity}
\label{apFP}

We want to prove that the parity operator defined in Eq.~(\ref{pk}) satisfies 
the conditions given by Eqs.~(\ref{P1}) and (\ref{P2}).

Inserting Eq.~(\ref{pk}) and its inverse, $$P_\kappa^{-1}=e^{-i\varphi_{\!_P}}\gamma^0\gamma^5 K_1,$$ 
into the left hand side of (\ref{P1}), and using that $(K_1)^2=1$,  
$\gamma^\mu\gamma^5=-\gamma^5\gamma^\mu$, $(\gamma^5)^2=\mathbb{1}$, and
$\gamma^0\gamma^\mu\gamma^0=\gamma^{\mu^\dagger}$ we get
\begin{equation}
P_\kappa^{-1}\gamma^\mu\Lambda_{\!\mu}^{\;\nu}P_\kappa 
=-\gamma^{\mu^\dagger}\Lambda_{\!\mu}^{\;\nu}.
\end{equation}
If we now use Eq.~(\ref{pmunu}) and that $\gamma^{0^\dagger}=\gamma^0$ and
$\gamma^{j^\dagger}=-\gamma^j$, we realize that 
$-\gamma^{\mu^\dagger}\Lambda_{\!\mu}^{\;\nu}=-\gamma^\nu$. Thus
\begin{equation}
P_\kappa^{-1}\gamma^\mu\Lambda_{\!\mu}^{\;\nu}P_\kappa 
=-\gamma^{\nu},
\label{f2a}
\end{equation}
which is equal to the right hand side of (\ref{P1}) if $\theta=\pi$.

Using that $\theta=\pi$ and multiplying the left and right hand sides of Eq.~(\ref{P2}) 
by $-mc$, we can recast it as follows,
\begin{equation}
P_\kappa^{-1} \overline{B}P_\kappa = - \overline{B},
\label{P2a}
\end{equation}
where $\overline{B} = -imc\gamma^5 + \hbar \kappa_\mu\gamma^\mu$. Inserting Eq.~(\ref{pk})
and its inverse into the left hand side of (\ref{P2a}), using the same identities for 
the $\gamma^\mu$ matrices just listed above, and employing Eq.~(\ref{k1}), we obtain
\begin{equation}
P_\kappa^{-1} \overline{B}P_\kappa = imc\gamma^5 - \hbar\kappa_\mu\gamma^\mu = 
-\overline{B}.
\label{f3}
\end{equation}
This is exactly the right hand side of (\ref{P2a}), proving 
consequently the validity of Eq.~(\ref{P2}). Note that the property 
(\ref{k1}) of the operator $K_1$ is crucial to arrive at Eq.~(\ref{f3}).

Equation (\ref{pk}) also implies that under the parity operation the 
four-current $j^\mu$ transforms according to what one expects of a 
contravariant four-vector. To see this, we repeat 
the argument given in Sec. \ref{cfc}. The
assumption that $j^\mu$ is a four-vector leads to
\begin{equation}
\gamma^\nu = [\gamma^0 P_\kappa^\dagger \gamma^0] 
\gamma^\mu \Lambda_{\!\mu}^{\;\nu} P_\kappa .
\label{cond0p}
\end{equation}
Comparing with Eq.~(\ref{f2a}) we see that we must have
\begin{equation}
\gamma^0 P_\kappa^\dagger \gamma^0 = - P_\kappa^{-1}
\label{f6}
\end{equation}
if $j^\mu$ transforms as a contravariant four-vector.
And a direct calculation, using the definition of $P_\kappa$, shows that 
Eq.~(\ref{f6}) is true.

Working in the Dirac-Pauli representation of the gamma matrices, we have that
the non-relativistic limit of the wave function of the asymmetric Dirac equation 
is given by Eq. (\ref{wfnrl}). A simple calculation using
$P_\kappa$ in this representation gives
\begin{equation}
P_\kappa \Psi_P(x) = ie^{i\varphi_{\!_P}}K_1\Psi_P(x),
\end{equation}
which tells us that the wave function for the asymmetric Dirac equation in the 
non-relativistic limit is an eigenvector of the parity operator. Moreover,
using Eqs.~(\ref{u0}) and (\ref{v0}) we have
\begin{eqnarray}
P_\kappa u_r(0) &=& ie^{i\varphi_{\!_P}}K_1 u_r(0), \label{f8}\\
P_\kappa v_r(0) &=& -ie^{i\varphi_{\!_P}}K_1 v_r(0). \label{f9}
\end{eqnarray}
Equations (\ref{f8}) and (\ref{f9}) show that in the particle's 
rest frame the spinors $u_r(0)$ and $v_r(0)$ have opposite parity, 
similarly to the behavior of the standard Dirac spinors $u_{\!_Dr}(0)$
and  $v_{\!_Dr}(0)$ \cite{gre00}.

Using that $\Psi'(x')=P_\kappa\Psi(x), 
\overline{\Psi'}(x')=-\overline{\Psi}(x)P_\kappa^\dagger,
P_\kappa^\dagger=P_\kappa^{-1}$, Eqs.~(\ref{si}), 
(\ref{pk}), (\ref{f2a}), (\ref{P2a}), and that
$[P_\kappa,\bm{\sigma}]=0$, we can show that 
\begin{eqnarray}
\mathcal{L'}(x') &=& \mathcal{L}(x), \label{lagP}\\
H'&=& H, \label{HP}\\
\mathbf{P'}&=& -\mathbf{P}, \label{PP}\\
\mathbf{J'}&=& \mathbf{J}, \label{JP}
\end{eqnarray}
where $\mathcal{L}$, $H$, $\mathbf{P}$, and $\mathbf{J}$ are
the Lagrangian density, the Hamiltonian, the 
linear momentum vector, and the total angular momentum vector related to
the asymmetric Dirac equation [cf. Eqs.~(\ref{adeld}), (\ref{ham}), (\ref{p}), 
and (\ref{jvec})]. Equations (\ref{lagP})-(\ref{JP}) are
the expected behavior of how $\mathcal{L}, H, \mathbf{P}$, and $\mathbf{J}$ 
transform after the space inversion operation, the same behavior one obtains
for the corresponding quantities related to the standard Dirac equation.

\subsection{Time reversal}

Our goal here is to prove that the time reversal operator (\ref{tk}) is
compatible with Eqs.~(\ref{T1}) and (\ref{T2}).

Inserting Eq.~(\ref{tk}) and its inverse, 
$T_\kappa^{-1}=T^{-1} K_1\gamma^5$, 
into the left hand side of Eq.~(\ref{T1}) leads to
\begin{equation}
T_\kappa^{-1}i\gamma^\mu\Lambda_{\!\mu}^{\;\nu}T_\kappa 
=i\gamma^{3}\gamma^{1}(K\gamma^\mu K\Lambda_{\!\mu}^{\;\nu})
(\gamma^1\gamma^3),
\end{equation}
where, in addition to the identities highlighted before 
in the Sec. 
\ref{apFP}, we
used that in the Dirac-Pauli representation $\gamma^2$ is an imaginary
matrix, i.e., $K\gamma^2K=-\gamma^2$, while for the real ones 
$K\gamma^\mu K=\gamma^\mu$, where
$\mu\neq 2$. Using these latter identities, the anticommutation properties 
of $\gamma^\mu$, and Eq.~(\ref{trmunu}) we get
\begin{equation}
i\gamma^{3}\gamma^{1}(K\gamma^\mu K\Lambda_{\!\mu}^{\;\nu})
(\gamma^1\gamma^3) = -i\gamma^\nu
\end{equation}
and thus
\begin{equation}
T_\kappa^{-1}i\gamma^\mu\Lambda_{\!\mu}^{\;\nu}T_\kappa 
=-i\gamma^\nu,
\label{f12}
\end{equation}
which proves the validity of Eq.~(\ref{T1}) if $\theta = \pi$.

Setting $\theta=\pi$ and multiplying the left and right hand sides of Eq.~(\ref{T2}) 
by $-mc$ we obtain the equivalent condition
\begin{equation}
T_\kappa^{-1} \overline{B}T_\kappa = - \overline{B},
\label{T2a}
\end{equation}
with $\overline{B}$ defined in 
\ref{apFP}.
Using Eq.~(\ref{tk}),
its inverse, the identities involving the gamma matrices highlighted in 
\ref{apFP},
and remembering that in the Dirac-Pauli representation $\gamma^5$ is real, the left hand side of (\ref{T2a}) becomes
\begin{equation}
T_\kappa^{-1} \overline{B}T_\kappa = imc\gamma^5 - \hbar\gamma^3\gamma^1
\left(\sum_\mu \kappa^\mu K\gamma^\mu K\right)\gamma^1\gamma^3.
\label{f14}
\end{equation}
However, a direct calculation gives
\begin{equation}
\gamma^3\gamma^1
\left(\sum_\mu \kappa^\mu K\gamma^\mu K\right)\gamma^1\gamma^3 
= \kappa^0\gamma^0- \sum_{j=1}^{3}\kappa^j\gamma^j = \kappa_\mu\gamma^\mu,
\end{equation}
and therefore Eq.~(\ref{f14}) becomes
\begin{equation}
T_\kappa^{-1} \overline{B}T_\kappa = imc\gamma^5 - \hbar\kappa_\mu\gamma^\mu = 
-\overline{B}.
\label{f16}
\end{equation}
This is equivalent to Eq.~(\ref{T2}) with $\theta=\pi$ that we wanted to prove
[cf. Eq.~(\ref{T2a})].

Equation (\ref{tk}) also gives that under the time reversal operation the 
four-current $j^\mu$ transforms according to a 
contravariant four-vector, namely, $j^0$ does not change its sign while
$j^i$ changes it. Repeating 
the calculations of Secs. \ref{cfc} and \ref{apFP}, and noting that 
now $ \Lambda_{\!\mu}^{\;\nu}$, Eq.~(\ref{trmunu}), is minus Eq.~(\ref{pmunu}), 
the assumption that $j^\mu$ is a four-vector gives
\begin{equation}
\gamma^\nu = -[\gamma^0 T_\kappa^\dagger \gamma^0] 
\gamma^\mu \Lambda_{\!\mu}^{\;\nu} T_\kappa .
\label{cond0t}
\end{equation}
Now, since $i$ anticommutes with $T_\kappa^{-1}$, Eq.~(\ref{f12}) becomes
\begin{equation}
T_\kappa^{-1}\gamma^\mu\Lambda_{\!\mu}^{\;\nu}T_\kappa 
=\gamma^\nu
\end{equation}
and a comparison with Eq.~(\ref{cond0t}) leads to
\begin{equation}
\gamma^0 T_\kappa^\dagger \gamma^0 = - T_\kappa^{-1}.
\label{f19}
\end{equation}
Finally, a direct calculation using the definition of $T_\kappa$ shows that 
Eq.~(\ref{f19}) is satisfied.

Employing that $\Psi'(x')=T_\kappa\Psi(x)$, 
$\overline{\Psi'}(x')=-\overline{\Psi}(x)T_\kappa^\dagger$,
$T_\kappa^\dagger=T_\kappa^{-1}$, Eqs.~(\ref{tr}), 
(\ref{tk}), (\ref{f12}), (\ref{f16}), 
$[T_\kappa,\gamma^0\bm{\kappa}]=0$, and that $[T_\kappa,\gamma^0\bm{\sigma}]=0$,
we obtain
\begin{eqnarray}
\mathcal{L'}(x') &=& \mathcal{L}(x), \label{lagT}\\
H'&=& H, \label{HT}\\
\mathbf{P'}&=& -\mathbf{P}, \label{PT}\\
\mathbf{J'}&=& -\mathbf{J}. \label{JT}
\end{eqnarray}
Equations (\ref{lagT})-(\ref{JT}) are
the expected behavior of how $\mathcal{L}, H, \mathbf{P}$, and $\mathbf{J}$ 
transform after the time reversal operation and they are similar to what one 
sees for the respective quantities associated with the standard Dirac equation.

We should also remark that the outcomes of the above calculations should be 
more rigorously written as, for instance, 
$\mathbf{J'}(x^{0'})= -\mathbf{J}(x^{0})$, where $x^{0}=-x^{0'}$. 
But since for a free
field we have that its energy, 
linear momentum, and total angular momentum are all conserved, 
we dropped the time dependency of the above expressions.

\subsection{Charge conjugation}

To prove Eqs.~(\ref{u1})-(\ref{u3}), we first note that 
if, without losing in generality, we set $\varphi_c=0$, 
Eqs.~(\ref{cg})-(\ref{cdirac}) give 
\begin{equation}
U_\kappa = U_\kappa^{-1}=i\gamma^2K_2.
\label{f20}
\end{equation}

Inserting Eq.~(\ref{f20}) into the left hand side of (\ref{u1}) we get
\begin{equation}
U_\kappa \gamma^{\mu^*} U_\kappa^{-1}  = 
-\gamma^2\gamma^{\mu^*}\gamma^2 = -\gamma^\mu,
\label{u1prova}
\end{equation}
where the latter equality is a consequence of the algebra of the gamma matrices and that 
we are working in the Dirac-Pauli representation, where $\gamma^2$ is a pure imaginary matrix while the other gamma matrices are real matrices. Equation (\ref{u1prova}) is 
exactly Eq.~(\ref{u1}) when $\theta=\pi$.

Similarly, if we use Eqs.~(\ref{f20}), (\ref{k2}), and remember that in the Dirac-Pauli 
representation for the gamma matrices $\gamma^2\gamma^{\mu^*}\gamma^2=\gamma^\mu$, the 
left hand side of Eq.~(\ref{u2}) becomes
\begin{equation}
U_\kappa \gamma^{\mu^*}\kappa_\mu U_\kappa^{-1}  =  \gamma^\mu \kappa_\mu, 
\label{u2prova}
\end{equation}
which is exactly the right hand side of Eq.~(\ref{u2}) if $\theta=\pi$.

Finally, the proof of Eq.~(\ref{u3}) is obtained by using the identity 
$\gamma^2\gamma^5\gamma^2=\gamma^5$ and that in 
the Dirac-Pauli representation of the gamma matrices $\gamma^5=\gamma^{5^*}$.
These facts, together with Eq.~(\ref{f20}), allow us to write the left hand side of 
Eq. (\ref{u3}) as
\begin{equation}
U_\kappa \gamma^{5^*}U_\kappa^{-1} = -\gamma^5,
\label{u3prova}
\end{equation}
which equals the right hand side of Eq.~(\ref{u3}) when $\theta=\pi$.

To obtain the equivalent expressions of Eqs.~(\ref{lagP})-(\ref{JP}) and 
(\ref{lagT})-(\ref{JT}), where $\Psi_c(x)$ is interpreted as the field describing
antiparticles, we have to go to the second quantization formalism and 
use the fermion normal ordering prescription in the 
calculations below \cite{rig23}. 
As such, if $\hat{\mathcal{O}}$ is any function of the gamma matrices not depending 
on fermion fields, we will have identities of the following type, 
\begin{displaymath}
\Psi^T\hat{\mathcal{O}}\overline{\Psi}^T =
-[\overline{\Psi}\hat{\mathcal{O}}^T\Psi]^T = -\overline{\Psi}\hat{\mathcal{O}}^T\Psi.
\end{displaymath}
The minus sign comes from the normal ordering prescription, i.e., whenever we exchange the order of two fermion operators we gain a minus sign. And the last equality above
follows from the fact that the transpose of a c-number is the c-number itself. 

With the above provisos,
using that $\Psi_c(x)=U_\kappa\Psi^*(x)$, 
$\overline{\Psi}_c(x') =-\Psi^T(x)\gamma^0 U_\kappa^\dagger$,
$U_\kappa^\dagger=U_\kappa^{-1}$, Eqs. (\ref{cg}), (\ref{u1prova}),
(\ref{u2prova}), (\ref{u3prova}), 
$\gamma^0\gamma^{\mu^*}\gamma^0=\gamma^{\mu^T}$, $[U_\kappa,\gamma^0\bm{\kappa}]=0$,
and that $U_\kappa^\dagger\gamma^0\bm{\sigma}U_\kappa=\bm{\sigma}^*\gamma^0$, 
we can show that 
\begin{eqnarray}
\mathcal{L}_c(x) &=& \mathcal{L}(x), \label{lagC}\\
\mathcal{L}_{c_{int}}(x) &=& -\mathcal{L}_{int}(x), \label{lagIC}\\
H_c&=& H, \label{HC}\\
\mathbf{P}_c&=& \mathbf{P}, \label{PC}\\
\mathbf{J}_c&=& \mathbf{J}, \label{JC}
\end{eqnarray}
where $\mathcal{L}, \mathcal{L}_{int}, H, \mathbf{P}$, and $\mathbf{J}$
are given respectively by Eqs. (\ref{adeld}), (\ref{adelint}), (\ref{ham}), (\ref{p}), 
and (\ref{jvec}). 

It is worth noting that during the calculations leading to 
Eq.~(\ref{lagC}) and to Eqs.~(\ref{HC})-(\ref{JC}), 
we obtain that the left hand sides and the right hand sides
are equal up to a four-divergence [Eq.~(\ref{lagC})] 
or to spatial volume integrals
that can all be converted to surface integrals that vanish since we are 
assuming the fields go to zero sufficiently fast as we tend to infinity.


\end{document}